\documentclass[twocolumn,showkeys,superscriptaddress]{revtex4-2}
\usepackage[utf8]{inputenc}
\usepackage{amsmath,amsfonts,amssymb,amsthm}
\usepackage{quotes}
\usepackage{graphicx}
\usepackage{hyperref}
\usepackage{cleveref}
\usepackage{xcolor}
\definecolor{goodblue}{RGB}{0, 91, 187}
\hypersetup{
  colorlinks=true,
  allcolors=goodblue,
  urlcolor=goodblue,
  citecolor=goodblue,
  pdfborder={0 0 0},
  breaklinks=true,
}
\usepackage[caption=false]{subfig}
\usepackage{booktabs}

\newcommand{\diag}{\text{diag}}

\newcommand{\xhdr}[1]{\vspace{0.0mm}\noindent{\textbf{#1.}}\hspace{0.5mm}}
\begin{document}
\title{Filtering higher-order datasets}

\author{Nicholas W. Landry}
\affiliation{Vermont Complex Systems Center, University of Vermont, Burlington, VT, 05405}
\affiliation{Department of Mathematics and Statistics, University of Vermont, Burlington, VT, 05405}
\affiliation{Department of Applied Mathematics, University of Colorado at Boulder, Boulder, CO, 80309}
\affiliation{Pacific Northwest National Laboratory, 1100 Dexter Ave N, Seattle, WA, 98109}
\email{nicholas.landry@uvm.edu}
\author{Ilya Amburg}
\affiliation{Pacific Northwest National Laboratory, 1100 Dexter Ave N, Seattle, WA, 98109}
\author{Mirah Shi}
\affiliation{Pacific Northwest National Laboratory, 1100 Dexter Ave N, Seattle, WA, 98109}
\affiliation{Department of Computer and Information Science, University of Pennsylvania, Philadelphia, PA, 19104}
\author{Sinan G. Aksoy}
\affiliation{Pacific Northwest National Laboratory, 1100 Dexter Ave N, Seattle, WA, 98109}

\begin{abstract}
Many complex systems often contain interactions between more than two nodes, known as \textit{higher-order interactions}, which can change the structure of these systems in significant ways. Researchers often assume that all interactions paint a consistent picture of a higher-order dataset's structure. In contrast, the connection patterns of individuals or entities in empirical systems are often stratified by interaction size. Ignoring this fact can aggregate connection patterns that exist only at certain scales of interaction. To isolate these scale-dependent patterns, we present an approach for analyzing higher-order datasets by filtering interactions by their size. We apply this framework to several empirical datasets from three domains to demonstrate that data practitioners can gain valuable information from this approach.
\end{abstract}


\maketitle

\section{\label{sec:introduction} Introduction}

Complex systems science provides a powerful framework for analyzing empirical systems because it not only accounts for individual entities but the interactions between them as well. These interactions shape the structure of complex systems at many scales~\cite{anderson_more_1972}. In these systems, one can quantify this structure by, for example, measuring the propensity of nodes with similar properties to connect with one another~\cite{newman_assortative_2002,newman_modularity_2006}, finding influential nodes~\cite{bonacich_power_1987,brin_anatomy_1998}, partitioning a system into different communities~\cite{newman_modularity_2006,peixoto_bayesian_2019}, and measuring the heterogeneity of connection~\cite{boguna_absence_2003}. Such structural properties can not only efficiently characterize a dataset~\cite{rosvall_information-theoretic_2007}, but can also inform its dynamical behavior such as contagion spread~\cite{pastor-satorras_epidemic_2015}, synchronization~\cite{restrepo_onset_2005}, and many other phenomena.

Pairwise networks, or graphs, represent complex systems as a collection of interactions involving only two entities. Common measures of pairwise networks include the degree and categorical assortativity~\cite{newman_mixing_2003,newman_assortative_2002}, community structure~\cite{girvan_community_2002}, clustering coefficient~\cite{watts_collective_1998}, and centrality~\cite{bonacich_power_1987}. Many empirical systems, however, contain not only pairwise interactions but interactions between more than two nodes, known as \textit{higher-order} interactions. Higher-order networks, also known as \textit{hypergraphs}, are the natural extension of pairwise networks~\cite{battiston_networks_2020} and can more accurately model the structure of higher-order empirical datasets~\cite{chodrow_configuration_2020}. Researchers have extended many notions of pairwise network structure to higher-order networks. These measures include degree assortativity~\cite{landry_hypergraph_2022,chodrow_configuration_2020}, categorical assortativity~\cite{chodrow_annotated_2020}, modularity~\cite{kaminski_clustering_2019}, community structure~\cite{chodrow_generative_2021}, centrality~\cite{benson_three_2019, tudisco_node_2021}, and degree heterogeneity~\cite{landry_effect_2020}, among others. Many of these higher-order measures extend metrics on pairwise networks through selection rules~\cite{chodrow_configuration_2020}, pairwise projections~\cite{benson_higher-order_2016,landry_hypergraph_2022}, or analyzing datasets with interactions all of the same size~\cite{benson_three_2019}.

Our central premise is that connection patterns are stratified by the interaction size. For this reason, subsets of the original higher-order dataset, restricted by their size, offer more granular insights into the network structure. These size-restricted subsets, called \textit{higher-order filterings}, are an effective tool for analyzing complex systems with higher-order interactions. Several studies assume that interactions of a certain size may be used to predict the interaction patterns of a different size~\cite{benson_simplicial_2018,young_hypergraph_2021} or that the dataset as a whole offers consistent information on the community structure~\cite{chodrow_generative_2021}, assortativity~\cite{chodrow_configuration_2020}, or other metrics. We assume this is not the case and observe that removing this assumption increases the information encoded in higher-order datasets. A similar idea has been explored for pairwise networks in Ref.~\cite{klein_emergence_2020}.
Recent work examining subsets of hypergraphs has examined degree-degree mixing~\cite{sun_higher-order_2021,landry_hypergraph_2022} and degree centrality~\cite{kovalenko_vector_2022}. In Ref.~\cite{lotito_higher-order_2022}, the authors count the number of motifs of different sizes to describe the structure of datasets.

Furthermore, our approach presents a compelling framework to analyze the introversion/extroversion of individuals in a higher-order interaction network. 
This question has been studied for pairwise networks in Ref.~\cite{newman_mixing_2003}, where the author associated introversion and extroversion with a node's degree and used degree assortativity to quantify how gregarious people interact. Higher-order interaction networks allow us to answer this question in a complementary way. Taking increasing interaction size as a proxy for increasing extroversion, we may filter the network into two parts: a part containing small-sized interactions and another part containing larger-sized ones. We then interpret individuals exhibiting large centralities among the smaller-sized interactions as strongly introverted and those with large centralities among the larger-sized interactions as strongly extroverted -- disregarding nodes that are very central in both regimes. By changing the threshold for "small"- and "large"-scale interactions, we glean nuanced insight into the introversion/extroversion properties of the network. In fact, our filtering framework allows us to examine a whole range of interaction types, and we could, for example, introduce a range of "intermediate" interaction sizes where central nodes are neither strong introverts nor strong extroverts.

In this work, we begin by formalizing our filtering approach in \Cref{sec:approach}. Next, in \Cref{sec:case_study}, we demonstrate its utility through a case study on an email dataset, where our framework allows us to examine unique associations, assortativity, centrality, and community structure across different filtering parameters. In the appendices, we develop further intuition for the filtering approach through analysis of synthetic datasets and then apply the framework to glean insights from real-life datasets across email, biology, and proximity domains. Our approach allows us to identify and study previously-obscured structure in the data.

\section{\label{sec:approach} The filtering approach}

Disaggregating interaction networks based on metadata is a common approach for pairwise networks, resulting in multiplex and multilayer networks. Common examples include separating transportation networks into layers by modality or company~\cite{gallotti_anatomy_2014,cardillo_modeling_2013}, social media platform~\cite{landry_limitations_2023}, or potential for transmission~\cite{adams_sex_2013}. The individual uniplex layers in these examples often have no intrinsic ordering and contain relatively few layers. In contrast to general multilayer networks, \textit{multislice networks} are composed of ordered layers~\cite{mucha_community_2010,kivela_multilayer_2014} such as variations across time~\cite{mucha_community_2010, mucha_communities_2010,bassett_dynamic_2011}, community structure of the same network at different scales~\cite{mucha_community_2010}, or network backbones with varying levels of sparsity~\cite{serrano_extracting_2009}.

From a higher-order perspective, several studies use subsets of higher-order data in their analysis. For example, Refs.~\cite{musciotto_detecting_2021,musciotto_identifying_2022} quantify significant interactions by measuring the over-expression of a given interaction with respect to the prediction of a null model, Ref.~\cite{st-onge_social_2021} examines the effect that removing interactions above a certain size has on disease dynamics, and Ref.~\cite{aksoy_hypernetwork_2020} measures how higher-order structure changes with respect to the minimum interaction overlap size, $s$.

In this paper, we define \textit{filtering} as an extension of the multislice approach for disaggregating network data. We present a general framework for more flexible analysis of higher-order datasets to better understand how structure is related to interaction size.

\begin{figure*}
    \centering
    \includegraphics[width=17.2cm]{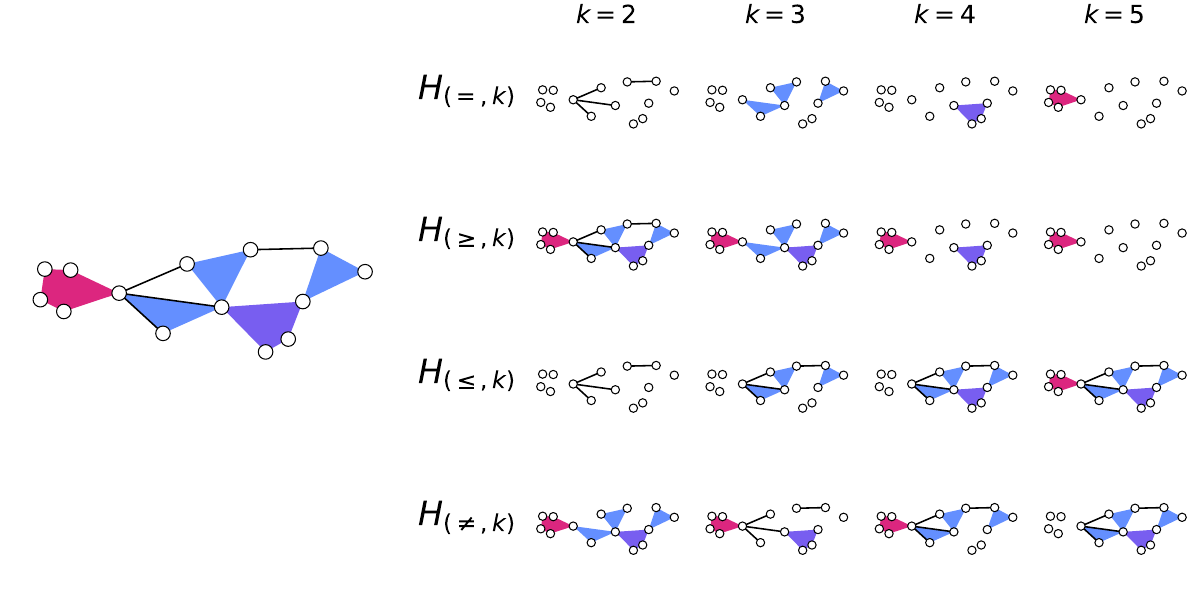}
    \caption{An illustration of a hypergraph with hyperedges of sizes two through five, visualized using XGI~\cite{landry_xgi_2023}. This hypergraph is decomposed using different types of filtering (uniform, $H_{(k,=)}$; GEQ, $H_{(k,\geq)}$; LEQ, $H_{(k,\leq)}$; and exclusion, $H_{(k,\neq)}$) for different filtering parameters, $k$.}
    \label{fig:viz}
\end{figure*}

\subsection{Empirical examples}

We present examples from social science, academia, and biology that illustrate how the scale of interactions can impact the underlying analysis.

\paragraph{Social networks}

The structure of social networks can be largely context-dependent. In multiplex networks, each uniplex layer formed by a different type of social tie may have significantly different structure~\cite{landry_limitations_2023} due to the various mechanisms of link formation~\cite{adams_sex_2013}. Studies have also shown that the extroversion of individuals affects their social network~\cite{zell_examining_2014,lepri_role_2016}. In Ref.~\cite{lepri_role_2016}, the authors show that "extroverts tend to have more complete triads and less incomplete or empty ones, than introverts." There is also evidence of a "rich club" phenomenon~\cite{zhou_rich-club_2004} where high-degree nodes connect with one another more often than would be expected at random. These pairwise network phenomena point to the possibility that the structure of social networks may change when examining interactions of different sizes.

\paragraph{Email} The etiquette of email communication often dictates whether an individual email will suffice or if a group email is appropriate and, if so, the number of people who should receive the message. In Ref.~\cite{roth_suggesting_2010}, the authors report that "group communication is so prevalent that our
analysis of the Google Mail email network shows that over 10\% of emails are sent to more than one recipient, and over 4\% of emails are sent to 5 or more recipients." These patterns are evident not only in the number of messages sent for different numbers of recipients, but also in the structure of the pairwise projection compared to the hypergraph as a whole~\cite{chodrow_annotated_2020}.

\paragraph{Co-authorship networks} Co-authorship networks are inherently higher-order; for example, a paper with three authors is not equivalent to three two-author papers. In general, the number of co-authors on a paper is a field-dependent distribution~\cite{newman_structure_2001}. Prior to the advent of higher-order network analysis, studies treated multi-author papers as cliques~\cite{palla_quantifying_2007}. More recent studies, however, have indicated that including these higher-order interactions can impact a co-authorship network's structure~\cite{chodrow_configuration_2020}. The authors of Ref.~\cite{patania_shape_2017} show that co-authorship networks largely obey triadic closure and that different scientific fields have different distributions of "maximal" collaborations.

\paragraph{Protein networks} Proteins interact with one another to form complex molecules that are essential to cell structure and function. The specific combination of these proteins can create very different resulting molecules, and combinations of more than two proteins can form a higher-order interaction network with vastly different topology than the equivalent pairwise network~\cite{klimm_hypergraphs_2021,murgas_hypergraph_2022}, suggesting that the structure of protein networks may vary significantly across interaction sizes.

\subsection{Terminology}

Consider a hypergraph $H = (V, E)$ where $V$ is the set of {\it nodes} and $E$ is the set of \textit{hyperedges}, which are arbitrary nonempty subsets of the node set. We call an interaction between $k$ entities a \textit{$k$-hyperedge}, and a hypergraph which solely consists of $k$-hyperedges is called \textit{$k$-uniform}. A \textit{simplicial complex} is a special case of a hypergraph where the existence of a $k$-hyperedge --- in this instance called a $(k-1)$-simplex --- implies the existence of every possible subset of that hyperedge.

We now define the \textit{filtering} of a hypergraph $H=(V,E)$ with respect to the \textit{filters} $f: V \to \{0,1\}$ and $g: E \to \{0,1\}$ to be the hypergraph $H_{f,g}=(V_f,E_g)$ where
\[
V_f=\{ v\in V \ | \ f(v) = 1\}
\]
is the set of filtered nodes with respect to $f$ and
\[
E_g = \{ e\in E \ | \ g(e) = 1\}
\]
is the set of filtered hyperedges with respect to $g$. For simplicity, we denote the set of nodes induced by the filtered edges as $V_g = \{v \ | \ v\in e\in E_g\}$. We note that the choice of $f$ must yield a node set $V_f$ such that $V_g \subseteq V_f$. Three common choices for the filtered set of nodes are (1) $f(v)=1$ and each filtering contains all the nodes, (2) $f(v) = \mathbb{I}_{V_f}(v)$, where $G_f=(V_f, E_f)$ is the giant component of the filtered edges $E_f$, and $\mathbb{I}$ is the indicator function, or (3), $f(v) = \mathbb{I}_{V_g}(v)$. In this paper, we solely consider the first choice to more easily compare nodal measures across different filterings. For ease of notation, we set $H_g \equiv H_{1,g}$.

We note that $g$ may be chosen to reflect a variety of network properties, as well as metadata associated with hyperedges. Both pairwise networks and hypergraphs may have metadata, however, so the benefit of using higher-order datasets is the representation of multiple interaction sizes. In addition, unlike categorical attributes, filtering edges by their size allows us to define an ordering on the hypergraph filterings. These reasons motivate us to solely consider filters based on hyperedge size.

For a given non-negative integer $k$ and a given comparison operator $*$, a \textit{size-dependent filtering} $H_{(*,k)}$ is a filtering $H_g$ where $g$ is defined by
\begin{equation}
g(e, k)=\begin{cases}1 &\mbox{ if } |e| * k \\
0 & \mbox{ otherwise}
\end{cases}.
\end{equation}
In the following, we define $k$ as the \textit{filtering parameter}.

More generally, one may define a size-dependent filtering based on a set $K$ of hyperedge cardinalities, but in this work, we consider a single value $k$. Below, we present examples of common comparison operators.

\begin{itemize}
    \item Uniform filtering, $H_{(=,k)}$. This type of filtering may be used to isolate the structure associated with a particular hyperedge size.
    \item GEQ filtering, $H_{(\geq, k)}$. This filtering can be used to see the effect of excluding pairwise interactions or hyperedges with smaller cardinality. One can call this the "higher-order" filtering for its ability to remove low-order interactions.
    \item LEQ filtering, $H_{(\leq, k)}$. This filtering can be used to exclude higher-order interactions. As before, one can also call this the "lower-order filtering." This has been used to construct some of the datasets in Ref.~\cite{benson_data_2021} (the author constructs hypergraphs with hyperedges of size $\leq 25$) and other papers that observe the effect of higher-order interactions by only including 2- and 3- hyperedges~\cite{grilli_higher-order_2017,ziegler_balanced_2022,iacopini_simplicial_2019}.
    \item Exclusion filtering, $H_{(\neq,m)}$. This filtering can be used to exclude interactions of a particular size. This can be helpful when considering datasets where a particular interaction size significantly alters the structure of a hypergraph --- whether through a large number of edges or a significantly anomalous structure --- and one desires to exclude that interaction size.
\end{itemize}

In principle, one can combine these operations to select or exclude more than one interaction size, but the filters presented in this paper form the basis for all other size-dependent filters.

We can generate a set of filterings by considering different values of $k$ or filtering operations. A set of filterings can either be \textit{disjoint} or \textit{overlapping}. A set of filterings, $\{H_{(*,k)} \ | \ k \in K\}$, where $K$ is a set of interaction sizes, is disjoint if $E_{(*, i)}\cap E_{(*, j)} = \varnothing$, for all distinct $i,j\in K$. An example of a disjoint filtering set is $\{H_{(=,k)} \ | \ k \in M\}$. A filtering set is overlapping when it is not disjoint. Examples of overlapping filtering sets are the complete sets of LEQ, GEQ, and exclusion filterings. Any hypergraph $H$ can be expressed as the union of all its filterings -- namely,
\begin{equation*}
H = \left(\bigcup_{k\in K}V_{f_k},\bigcup_{k\in K}E_{f_k}\right).
\end{equation*}
For the uniform filtering set, the edge-wise union is disjoint. When filtering simplicial complexes, only the LEQ filtering preserves the simplicial structure of the dataset, but one could, in principle, convert the simplicial complex to its equivalent hypergraph to allow the use of other filtering types.

\subsection{Disconnected filterings}

It is common that despite the original dataset being fully connected, the filtered dataset is not, and we can deal with this in several ways.

First, we note that many higher-order measures are not dependent on the dataset being fully connected. For example, degree assortativity and modularity are aggregated over nodes and edges without any dependence on the connectedness of the dataset. Similarly, the degree of a node simply measures the number of hyperedges of which that node is a member.

This is not always the case, however. Several measures of centrality~\cite{benson_three_2019,tudisco_node_2021} require the dataset to be connected. These eigenvector-based methods require a single connected component, although one can look at each connected component individually and determine the relative centrality of each node in that component. Alternatively, one can consider a regularized version of that metric, just as PageRank \cite{brin_anatomy_1998} is similar to eigenvector centrality and does not require the network to be connected. Likewise, in the case of community detection, when the dataset is not fully connected, there are several well-established ways to label isolated nodes and nodes not contained within the giant component~\cite{peixoto_bayesian_2019}.

\section{\label{sec:case_study} A case study}

\begin{figure*}
    \centering
    \begin{tabular}{cc}
        \subfloat[Effective information \label{fig:effective_information_email-enron}]{\includegraphics[width=7.8cm]{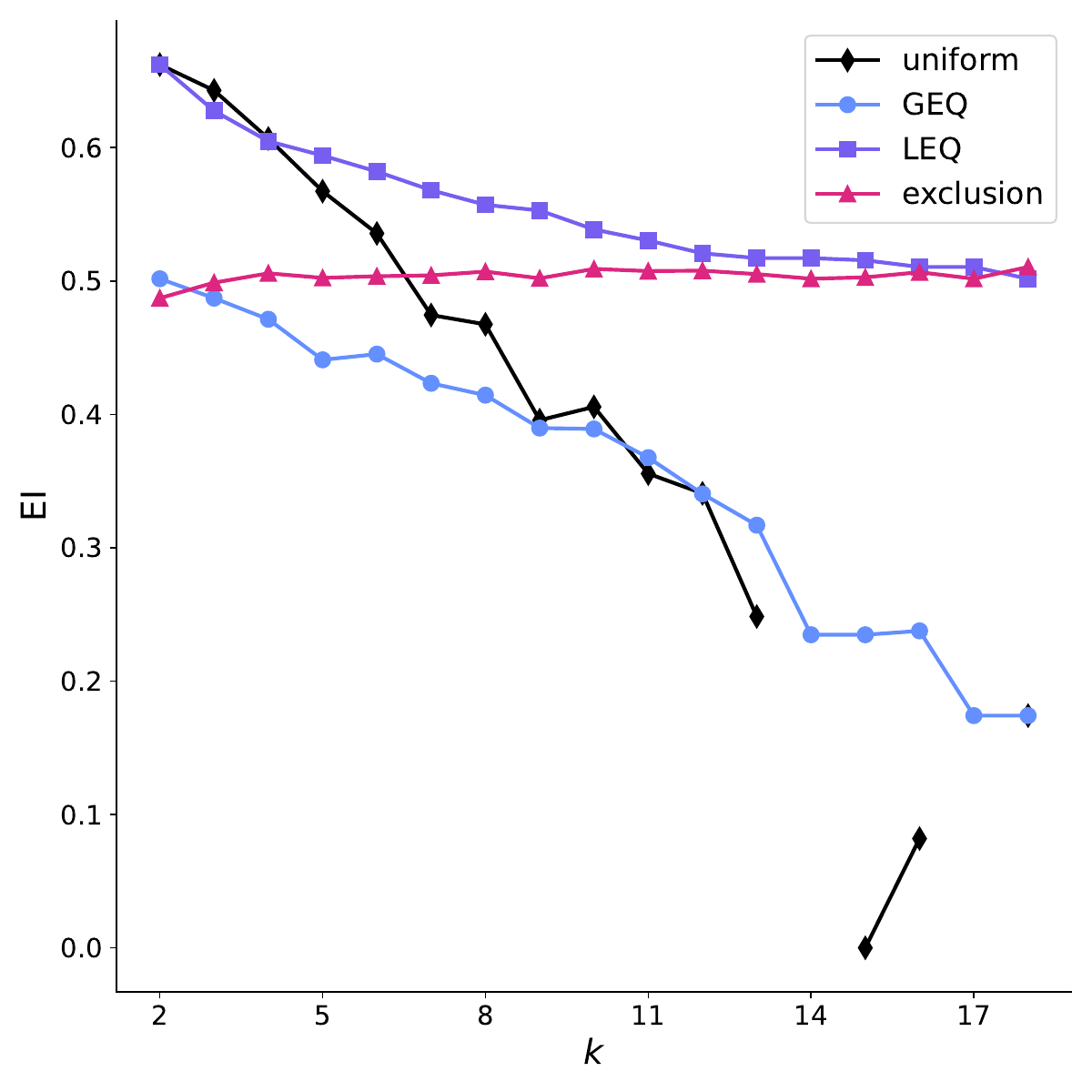}} &  \subfloat[Degree assortativity \label{fig:assortativity_email-enron}]{\includegraphics[width=8cm]{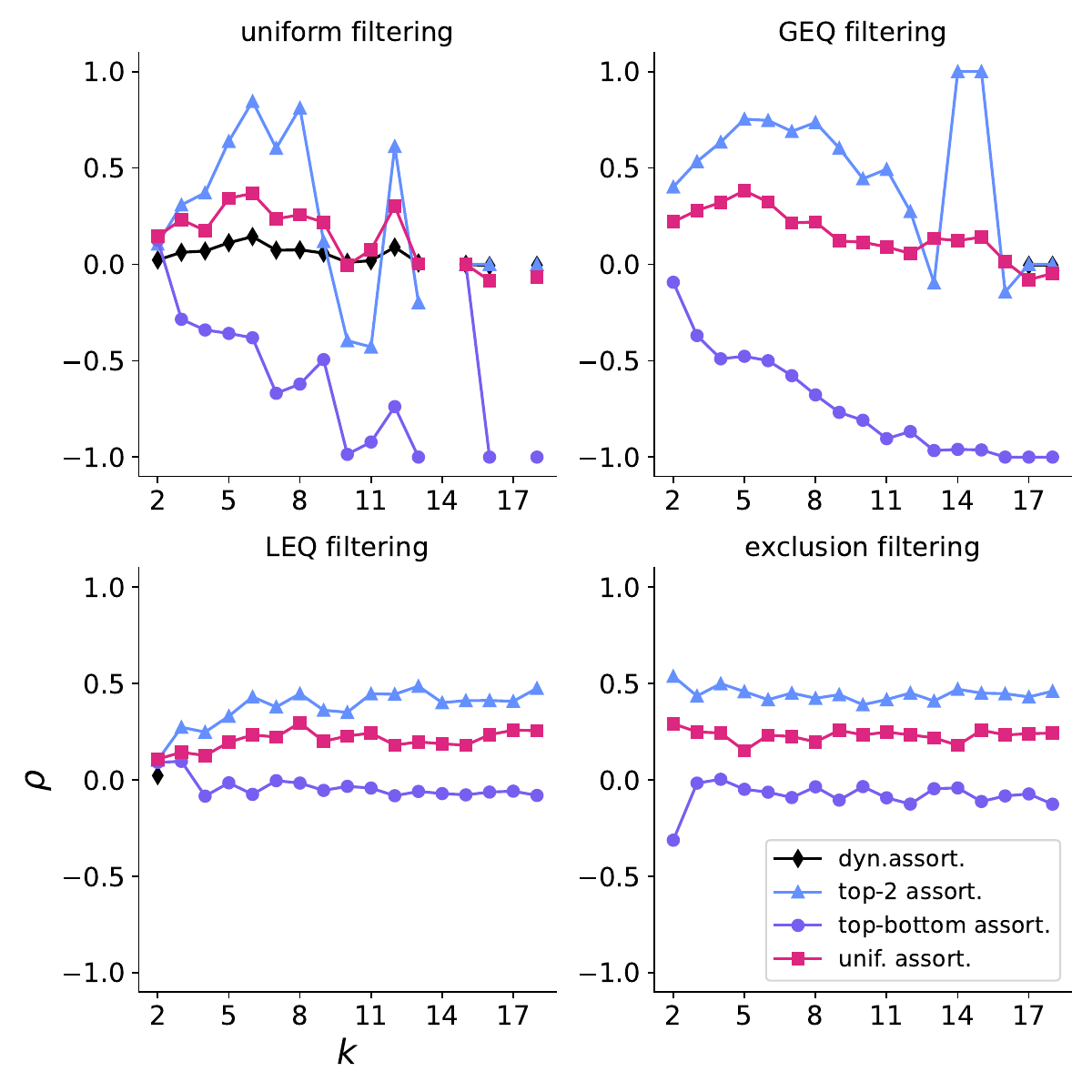}}
    \end{tabular}
    \caption{Global higher-order structural measures with respect to the filtering parameter $k$.}
    \label{fig:global_measures_email-enron}
\end{figure*}

In this section, we filter a single higher-order dataset to demonstrate how the structure of this dataset changes with interaction size. We present results for additional empirical datasets in \Cref{sec:additional_datasets}, where we also present intuitive filtering results on synthetic datasets. Those results demonstrate the utility of our approach across various domains.

\subsection{Measures of higher-order structure}

Here, we briefly describe the structural measures we utilize in this study and refer the reader to \Cref{sec:measures} for additional details.

\xhdr{Global measures} We consider the effective information and the degree assortativity as global measures of higher-order structure.
\begin{itemize}
\item Effective information (EI) is defined on the clique projection of the hypergraph as
\begin{equation*}
EI = \mathcal{H}(\langle W^{out}_i\rangle) - \langle \mathcal{H}(W^{out}_i)\rangle,
\end{equation*}
where $W^{out}_i$ is the $\ell_1$-normalized vector of out-degrees for node $i$ and $\mathcal{H}$ is the Shannon entropy. EI measures the strength of unique associations in a network.

\item Degree assortativity measures the tendency of nodes with similar degrees to connect with one another more often than would be expected at random. We utilized four different measures of degree assortativity; the first three --- top-bottom, top-2, and uniform --- defined as in Ref.~\cite{chodrow_configuration_2020}, and the dynamical assortativity~\cite{landry_hypergraph_2022} defined as
\begin{equation*}
\rho = \frac{\langle k\rangle^2\langle k k_1\rangle_E}{\langle k^2\rangle^2} - 1,
\end{equation*}
where $\langle k^r\rangle$ is the $r$-moment of the degree and $\langle k k_1\rangle_E$ is the expected pairwise product of degrees over hyperedges.
\end{itemize}

\xhdr{Local measures} We consider the betweenness centrality and the community labels of nodes as local measures of higher-order structure.

\begin{itemize}
\item
The node-based analogue of betweenness centrality for hypergraphs defined in Ref.~\cite{feng_hypergraph_2021} is
\begin{equation*}
BC(n) = \sum_{u \neq v \neq n\in V}\frac{\sigma_{uv}(n)}{\sigma_{uv}},
\end{equation*}
where $\sigma_{uv}$ is the number of shortest paths from node $u$ to node $v$ and $\sigma_{uv}(n)$ is the number of these shortest paths that pass through node $n$. Central nodes via this measure serve as "bridge" nodes between different parts of a network.

\item Community structure describes the mesoscale structure of complex systems by assigning the same labels to nodes that are densely connected with one another. To infer the community labels of nodes, we perform spectral clustering on the normalized Laplacian of the hypergraph, as proposed in Ref.~\cite{zhou_learning_2006}. We use a Hungarian matching algorithm to heuristically match two different sets of community labels.
\end{itemize}

\subsection{Filtering the email-enron dataset}

Here, we analyze the email-enron dataset, a higher-order dataset generated from emails sent from a core set of employees at Enron~\cite{landry_xgi-data_2023,cohen_enron_2015,benson_simplicial_2018}, where nodes represent email addresses and edges represent email messages. Before analyzing this dataset, we removed isolated nodes, multi-edges, and singleton edges. We comment that our results are qualitatively similar to those computed when including these artifacts. The resulting dataset has 143 nodes and 1457 edges and has heterogeneous degree and edge size distributions.

We analyze this dataset using the four types of filtering that we introduced: the uniform, LEQ, GEQ, and exclusion filterings. This provides a broad overview of the different types of filtering that can be used to examine a dataset. The structural metrics that we consider are by no means exhaustive but provide an instructive example of how structure can change for different interaction sizes. To illustrate our filtering approach, we present the effective information, degree assortativity, and inferred community structure in this section. The interaction size affects not only the local structure of a higher-order network but the global structure as well. We start by analyzing the global structure presented in \Cref{fig:global_measures_email-enron}.

\begin{figure*}
    \centering
    \begin{tabular}{cc}
        \subfloat[The $\ell_\infty$-normalized betweenness centrality of each node. Disconnected nodes have a centrality of zero by definition. \label{fig:betweenness_centrality_email-enron}]{\includegraphics[width=8cm]{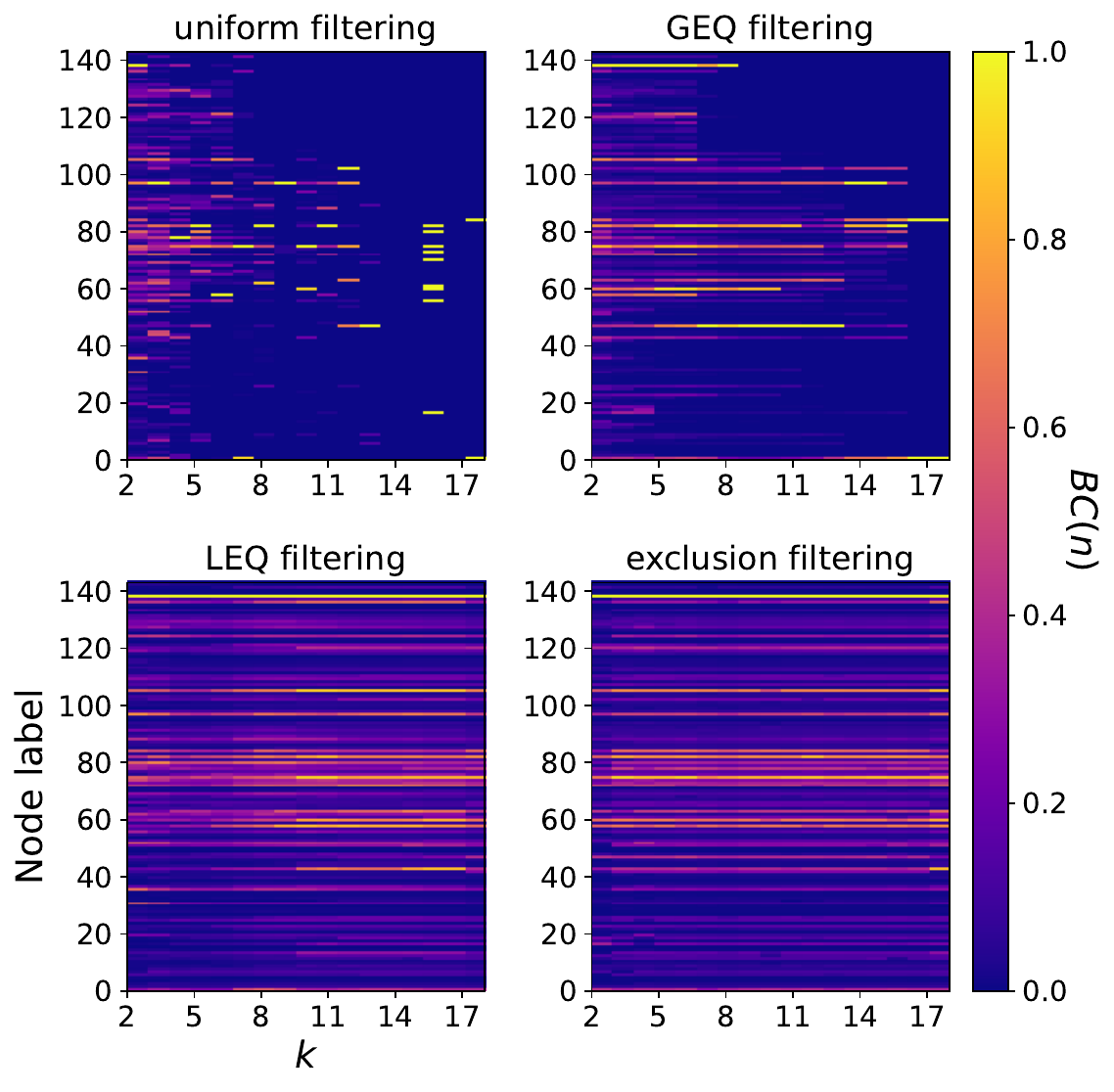}} &  \subfloat[The community labels of each node. White regions indicate isolated nodes. \label{fig:community_labels_email-enron}]{\includegraphics[width=8cm]{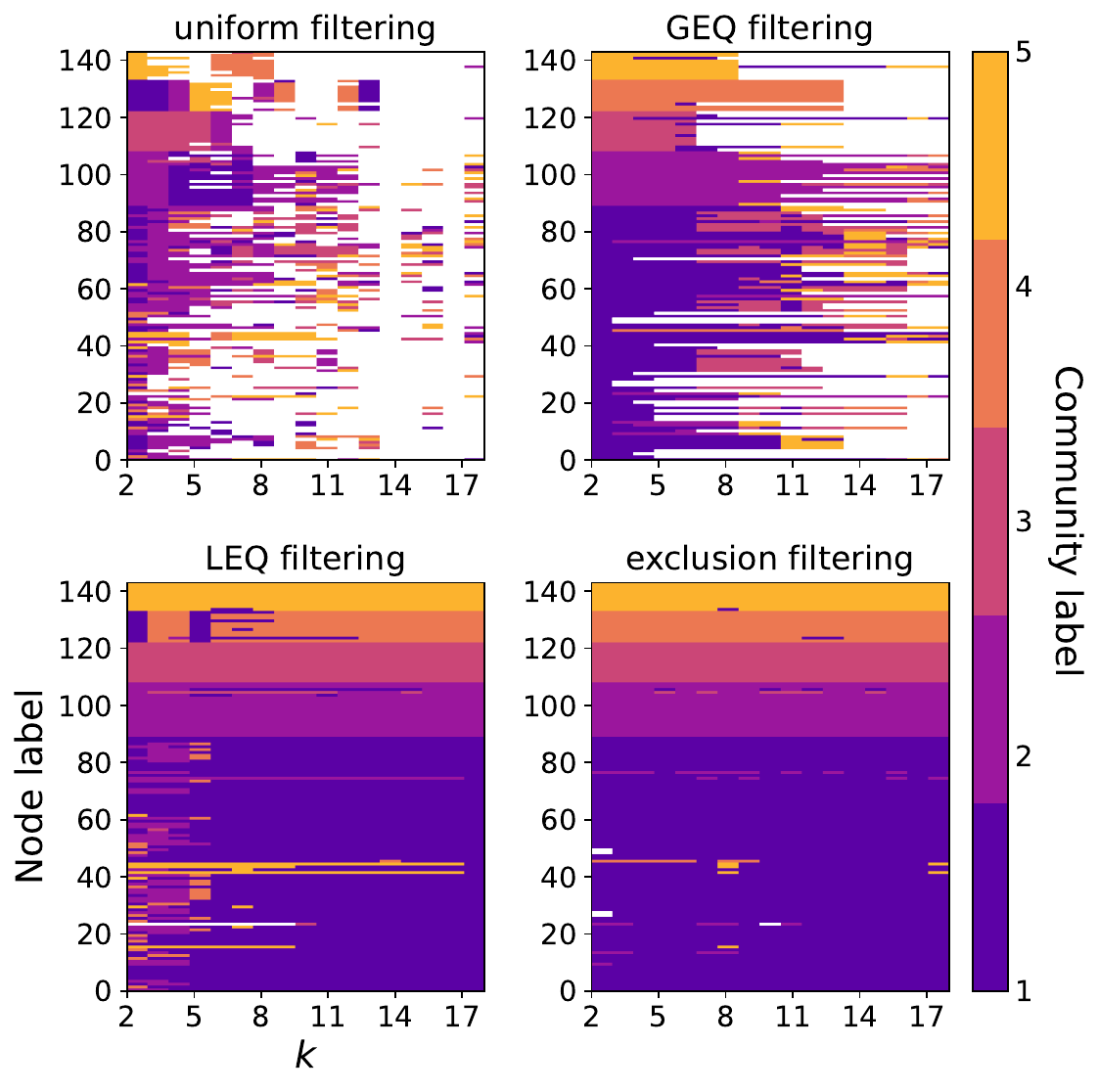}}
    \end{tabular}
    \caption{Nodal higher-order structural measures with respect to the filtering parameter $k$.}
    \label{fig:nodal_measures}
\end{figure*}

As seen in \Cref{fig:effective_information_email-enron}, 
the trends in normalized effective information show substantial differences across the uniform, GEQ, LEQ, and exclusion filterings. While the normalized effective information largely decreases with $k$ for the uniform, GEQ, and LEQ filterings, it largely \textit{increases} for the exclusion filtering. However, these trends are not monotonic; for example, the normalized effective information for the GEQ filtering increases from $k=14$ through $k=16,$ while that for the exclusion filtering decreases from $k=16$ to $k=17.$ Note that the effective information is not defined for $k=14, 17, 18$ for the uniform filtering since no hyperedges of those sizes appear in the dataset.

In \Cref{fig:assortativity_email-enron}, we see that for both the uniform and GEQ filtering, the degree assortativity increases with interaction size before declining again. For small values of $k$, the assortativity fluctuates less for the GEQ filtering when compared with the uniform filtering because there are far more edges over which to average. In addition, the exclusion filtering illustrates that the pairwise interactions most significantly affect the degree assortativity. Lastly, the LEQ filtering demonstrates that we can capture most of the assortative structure with interactions of size ten and smaller.

In \Cref{fig:betweenness_centrality_email-enron}, we see that the betweenness centrality of all nodes is sensitive to the filtering parameter. Comparing the LEQ and GEQ filterings, we see that the centralities are much more stable when excluding higher-order interactions compared with excluding lower-order interactions. Furthermore, these centrality results provide fertile ground for data insights. For example, Bill Williams (node 137), an Enron trader, has the largest centrality in the uniform filtering for $k=2$ but relatively small centralities for all other parameters. It turns out that he participated in many pairwise interactions (emails involving just one sender and recipient). Interestingly, his betweenness centrality remains high for a much larger parameter range for GEQ, LEQ, and exclusion filterings, indicating that he also serves as an intermediary for higher-order information. Another interesting trend is the very high betweenness centrality of Stanley Horton (node 46), President and CEO of Enron Transportation Services, in the GEQ filtering for intermediate values of $k$, indicating that he serves as a vital conduit of information in groups of intermediate size. In the other filterings, Stanley's centrality looks rather unremarkable across all filtering parameters.

The community structure presented in \Cref{fig:community_labels_email-enron} demonstrates that the assigned communities can change with differing interaction sizes. From the LEQ filtering, it seems that there are groups of nodes that switch memberships as larger interaction sizes are included. For the GEQ filtering, not only do we see nodes fail to join groups of a given size or larger, but their participation in a given community changes as well. In addition, the exclusion filtering indicates that interactions of a single size can be enough to change the community to which certain nodes belong. In contrast to existing literature suggesting that each node in a complex system has a single community label, relaxing this assumption leads to interesting trends in the community structure for different scales of interaction.

This case study illustrates the utility of filtering higher-order datasets by interaction size. Relaxing the assumption that all interactions contribute to a unified picture of a dataset can reveal structure that only exists at a particular scale. For all four measures of structure, we see noticeable changes when increasing the filtering parameter. Such changes can uncover the influence of different interaction sizes on the complete dataset.

\subsection{Gleaning insight across data domains}

In the appendices, we present multiple case studies across email, biological, and proximity domains. In the email and biological datasets, measuring effective information allows us to observe that higher-order interactions catalyze fewer unique associations than lower-order ones. Within the email datasets, taking interaction size as a proxy for gregariousness, we can identify key players at precise levels on the introvert/extrovert interaction scale. Stanley Horton stands as a compelling example from our Enron analysis, with his importance peaking within interactions of intermediate size. Further, among all the domains examined, community structure remains largely unchanged across all filterings in only the proximity datasets, with other domains showing complex changes in community structure with the filtering parameter.

\section{\label{sec:discussion} Discussion}

We have presented a framework for looking at subsets of higher-order datasets and demonstrated its usefulness by examining an empirical case study and offering insights from datasets across multiple domains. In particular, we used our filtering framework to study global properties like the strength of unique associations and assortativity, as well as local properties such as centrality and community structure. By examining centrality at different scales of interaction, we offered an approach for identifying introverts and extroverts in a population~\cite{newman_mixing_2003} by looking at the sizes of interactions in which they participate instead of simply their pairwise network degree. We believe that filtering higher-order datasets by interaction size is a valuable tool that reveals the stratification of connection patterns at different scales. There is still much to study on the theory and practice of filtering datasets; we have merely introduced it as a tool for the network science practitioner. Other important questions remain, however. When is it practical and helpful to look at subsets of datasets? Are there heuristics for deciding not only whether to filter a dataset but also the type of filter to use? Can we quantify the information gained by no longer constraining, for example, nodes to have a single community label when performing community detection?

When using this approach, caution is advised. By construction, filtering a higher-order dataset will yield fewer interactions, making the resulting network more susceptible to noise. This can be combatted by choosing sufficiently large or dense datasets. Quantifying the statistical significance and stability of different metrics for different filterings and datasets may be a fruitful future direction. Another pitfall is that the sparsity of a dataset's filterings can drive the observed structural changes, warranting more study to measure and correct for these effects.

Despite these limitations, filtering higher-order datasets by interaction size is a powerful approach and further sheds light on the assumptions made when quantifying complex systems. The consistency in the structural information that different filterings provide is a spectrum where, on one extreme, different filterings offer no information about each other, and on the other extreme, different filterings provide perfect information about each other. This notion of structural consistency across subsets of a dataset determines, for example, how similar the community labels are across all size-dependent filterings. Often, the default approach assumes perfect structural consistency, which we have demonstrated is not always the case. It may prove useful to quantify the information that two filterings of a higher-order dataset share to determine how much the structure of a dataset varies with interaction size.

This approach offers a counterpoint to the complex systems literature: in this case, the sum of the parts may be greater than that of the whole. This method of disaggregation allows us to observe how contact patterns can be stratified by interaction size. This has implications for sub-fields of network science, such as community detection, dynamics on networks, and structural measures, among other topics. We believe that our approach unifies studies that have indirectly examined the effect of size on structure and dynamics of higher-order datasets~\cite{chodrow_configuration_2020,kovalenko_vector_2022,musciotto_detecting_2021,musciotto_identifying_2022,st-onge_social_2021,sun_higher-order_2021} and will be a fruitful area of research in the future.

\section*{Data and Software} All software used to generate the results in this paper is available on \href{https://github.com/nwlandry/filtering-higher-order-datasets}{GitHub} \cite{landry_code_2023} and utilizes the XGI library~\cite{landry_xgi_2023}. All datasets used are openly available in the \href{https://gitlab.com/complexgroupinteractions/xgi-data}{XGI-DATA} repository.

\section*{Contributions}
N.W.L., I.A, M.S., and S.G.A. designed the research; N.W.L. and I.A. performed the research; N.W.L. and I.A. wrote the article; N.W.L., I.A., and S.G.A. edited the draft; and S.G.A. provided supervision.

\acknowledgments{N.W.L. acknowledges financial support from the National Science Foundation Grant 2121905, "HNDS-I: Using Hypergraphs to Study Spreading Processes in Complex Social Networks," and from the National Institutes of Health 1P20 GM125498-01 Centers of Biomedical Research Excellence Award. PNNL Information Release Number: PNNL-SA-184779. Pacific Northwest National Laboratory is operated by Battelle Memorial Institute under Contract DE-ACO6-76RL01830.}

\appendix
\onecolumngrid

\section{Structural measures\label{sec:measures}}

In this paper, we explore two types of structural measures: global and nodal. These measures are not exhaustive but are intended to give a large-scale picture of how the structure of complex systems changes with respect to interaction size.

\subsection{Global measures}
We consider the effective information and degree assortativity as global measures of higher-order structure. Effective information was introduced in Ref.~\cite{klein_emergence_2020} for directed graphs and is defined as $EI = \mathcal{H}(\langle W^{out}_i\rangle) - \langle \mathcal{H}(W^{out}_i)\rangle$, where $W^{out}_i$ is the $\ell_1$-normalized vector of out-degrees for node $i$ and $\mathcal{H}$ is the Shannon entropy. We extend this definition to hypergraphs by performing the degree computations on the clique projection associated with the weighted adjacency matrix $A = BB^T$ (setting the diagonal elements to zero), where $B$ is the incidence matrix. The authors of Ref.~\cite{klein_emergence_2020} explain this metric as quantifying the strength of unique associations in a network.

Degree assortativity, the preferential tendency of nodes with similar degrees to connect with one another, has been extended to hypergraphs in several ways. In Ref.~\cite{chodrow_configuration_2020}, the author defines the top-bottom, top-2, and uniform degree assortativity for non-uniform hypergraphs by selecting two nodes from a given hyperedge according to a given rule (the nodes with the smallest and largest degrees, the nodes with the two largest degrees, and two randomly selected nodes) and then computing the Pearson correlation coefficient. In our case, we approximate this metric by computing the assortativity for a sample of $10^3$ hyperedges.

In contrast to this method, Ref.~\cite{landry_hypergraph_2022} derives the dynamical assortativity for uniform hypergraphs with respect to a mean-field approximation of the largest eigenvalue as
\begin{equation*}
\rho = \frac{\langle k\rangle^2\langle k k_1\rangle_E}{\langle k^2\rangle^2} - 1,
\end{equation*}
where $\langle k^r\rangle$ is the $r$-moment of the degree and $\langle k k_1\rangle_E$ is the mean pairwise product of degrees over all possible 2-node combinations in each hyperedge in the hypergraph. We note that because Ref.~\cite{landry_hypergraph_2022} only defines dynamical assortativity for uniform hypergraphs, we can only compute it for the uniform filtering case (or whenever the filtering contains hyperedges of only one size, as is always the case for the largest parameter in GEQ).

\subsection{Nodal measures}
The nodal structural properties that we consider are the centrality and community labels of nodes. There are many notions of centrality for hypergraphs~\cite{benson_three_2019,tudisco_node_2021,feng_hypergraph_2021}. In this study, we use a node-based analogue of the betweenness centrality for hypergraphs defined in Ref.~\cite{feng_hypergraph_2021} as
\begin{equation*}
BC(n) = \sum_{u \neq v \neq n\in V}\frac{\sigma_{uv}(n)}{\sigma_{uv}},
\end{equation*}
where $\sigma_{uv}$ is the number of shortest paths from node $u$ to node $v$ and $\sigma_{uv}(n)$ is the number of these shortest paths that pass through node $n$. Here, the number of shortest paths is computed on the unweighted clique projection. In all experiments, we report the $\ell_{\infty}$-normalized betweenness centrality to facilitate easy comparisons across filterings and filtering parameters.

There are many algorithms for community detection on hypergraphs~\cite{zhou_learning_2006,kaminski_clustering_2019,chodrow_generative_2021,chodrow_nonbacktracking_2023}, and for this study, we use the method proposed in Ref.~\cite{zhou_learning_2006}. We start by defining the normalized Laplacian of the hypergraph as
\begin{equation*}
L = \frac{1}{2}\left(I - D^{-1/2} A D^{-1/2}\right),
\end{equation*}
where $A$ is defined as before and $D=\diag(A{\bf 1})$ is the diagonal matrix of degrees. We then compute the $l$ largest eigenvectors of this matrix and perform $k$-means clustering on these vectors. One caveat is that the number of clusters is fixed to $l$, regardless of the number of connected nodes. This can sometimes lead to well-defined clusters being split into two or more clusters or other changes in cluster assignments.

\subsubsection{Cluster label matching}

To help visualize communities, we reorder nodes so that all nodes in the same community in the GEQ filtering for $k=2$ appear together, and the largest communities appear on the bottom. To compare the communities output by the clustering algorithm across filtering parameters within a size-dependent filtering, we perform a weighted Jaccard-index-based Hungarian matching~\cite{kuhn_hungarian_1955} to sequentially match current communities to previous communities as we sweep across the parameter. To compare communities across filterings, we reorder the nodes based on the ordering induced by the GEQ filtering and then match communities for $k=2$. We then match across filtering parameters within the filtering in question as described above.

\section{Results from synthetic hypergraphs}

For each of the metrics we have described --- effective information, degree assortativity, community structure, and centrality --- we describe synthetic models of higher-order networks for which we can analytically compute the structure or better describe why the structure changes with interaction size.

\subsection{Effective information}
\begin{figure}[h]
    \centering
    \includegraphics{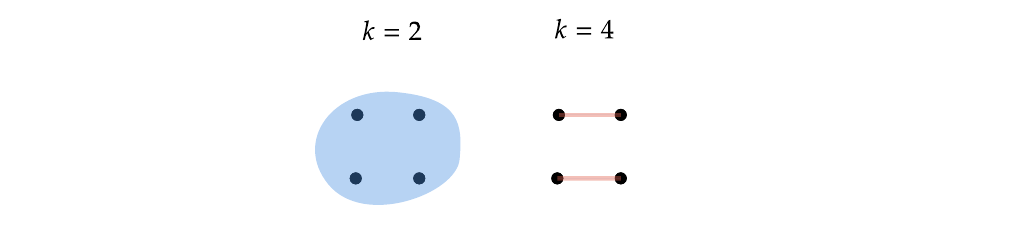}
    \caption{(Left) Exclusion filtering for $k=2$. (Right). Exclusion filtering for $k=4$. We observe that the EI increases from $1 - \log(3)/\log(4)$ to 1 as we increase the filtering parameter $k$ from 2 to 4.}
    \label{fig:toy_effective_info}
\end{figure}

Consider a hypergraph with four nodes, two edges linking two disjoint pairs of nodes, and a hyperedge including all nodes. The exclusion filtering of the hypergraph with $k=2$ and $k=4$ is shown in Figure~\ref{fig:toy_effective_info}. It is easy to verify that the hypergraph corresponding to $k=2$ has a normalized effective information of $1 - \log(3)/\log(4),$ while that corresponding to $k=4$ has a higher normalized EI of 1.

\newpage

\subsection{Degree assortativity}
For this metric, we use the sunflower hypergraph model as presented in Ref.~\cite{benson_three_2019} where we select the number of petals, $n_p$, the interaction size, $k$, and the number of nodes, $c$ to be members of every petal. We populate each petal with the $c$ center nodes and $k - c$ isolated nodes. The resulting hypergraph has $n_p$ edges and $(k - c)n_p + c$ nodes. We compute the terms in the expression for dynamical assortativity for each edge size $k$ as follows:
\begin{align*}
\langle d \rangle &= \frac{k n_p}{(k - c)n_p + c},\\
\langle d^2\rangle &= \frac{(k - c) n_p + c n_p^2}{(k - c)n_p + c},\\
\langle d d_1\rangle_E &= \frac{1}{\binom{k}{2}}\sum_{i<j} d_i d_j\\
&= \frac{1}{\binom{k}{2}}\left[ \binom{c}{2}n_p^2 + n_p(k - c)c + \binom{k-c}{2} \right],
\end{align*}
and simplifying the final expression, we obtain
\begin{equation}
\rho = -\frac{c(k-c)(n_p - 1)^2}{(k - 1)[k +c(n_p - 1)]^2}.
\end{equation}

\begin{figure*}[h]
    \centering
    \includegraphics[width=\textwidth]{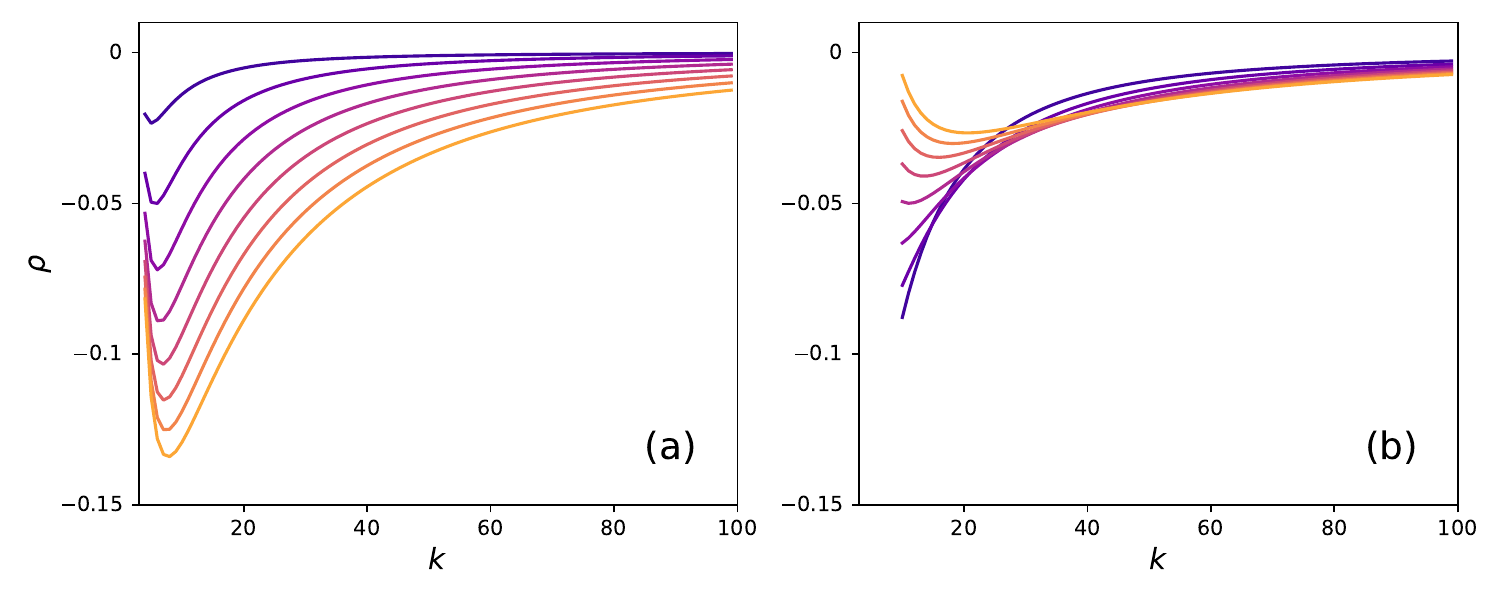}
    \caption{A plot of the dynamical assortativity of a sunflower hypergraph for (a) different number of petals, $n_p$ and (b) different size centers, $c$. The line colors range from dark to light as the number of petals in panel (a) and the size of the sunflower center in panel (b) are increased from two to nine. In panel (a), the center size is fixed at $c=3$, and in panel (b), the number of petals is fixed at $n_p=5$.}
    \label{fig:toy_assortativity}
\end{figure*}
From this expression, we can see that the hypergraph will always be disassortative because $k > c \geq 1$, and as the edge size increases, the low-degree nodes will "dilute" the assortativity, eventually leading to a hypergraph with 0 assortativity. ee \Cref{fig:toy_assortativity} for a visualization of this trend.

\newpage

\subsection{Centrality}
\begin{figure}[h]
    \centering
    \includegraphics[width=0.59\linewidth]{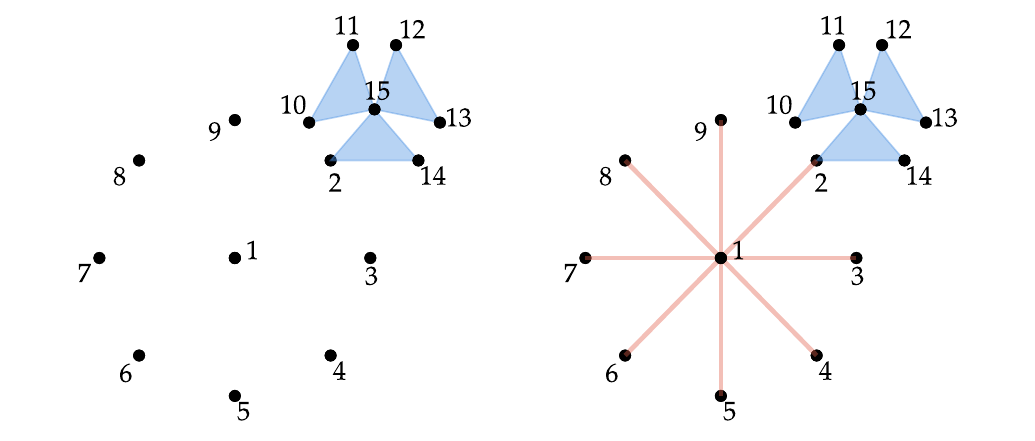}
    \caption{(Left). LEQ filtering for $k= $. (Righ ). LEQ filtering for $k=3$ with communities superimposed.}
    \label{fig:toy_centrality}
\end{figure}

Consider a star graph with eight edges joined together at a node with a sunflower having $n_p=3$ and $k=3$, as seen in Figure~\ref{fig:toy_centrality}. n the subhypergraph generated by the LEQ filtering for $k=2,$ node 15 has an $\ell_\infty$-normalized betweenness centrality of 1 (and the rest 0). However, this drastically changes in the subhypergraph corresponding to $k=3$, as the normalized centrality of node 15 decreases to 11/14, that of node 1 increases to 1, and that of node 2 increases to 24/35. Node 1 transitions from being the least to the most central node in the network.

\subsection{Community structure}
\begin{figure}[h]
    \centering
    \includegraphics[width=0.59\linewidth]{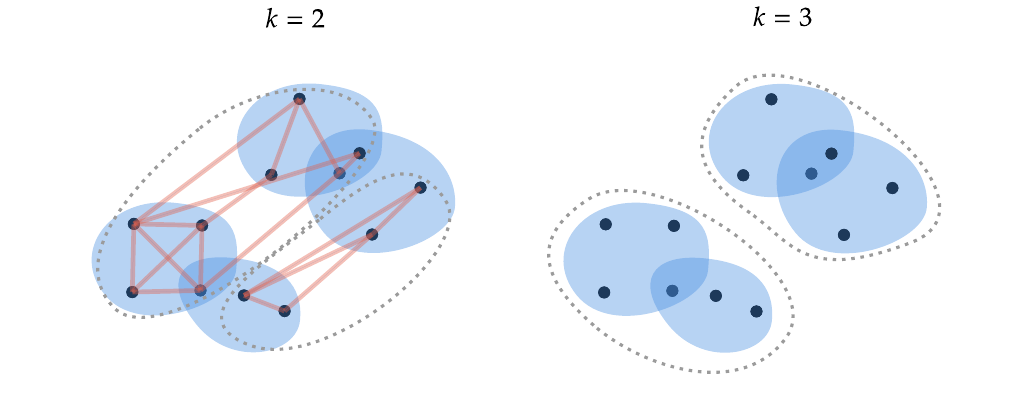}
    \caption{(Left). GEQ filtering for $k=2$ with communities (dotted) superimposed. (Right). GEQ filtering for $k=3$ with communities superimposed.}
    \label{fig:toy_communities}
\end{figure}

The hypergraph shown in Figure~\ref{fig:toy_communities} contains two dense communities (left, surrounded by dotted ovals), but when applying the GEQ filtering for $k=3$ (right), two completely different communities emerge.

\newpage

\section{\label{sec:additional_datasets} Additional datasets}

\subsection{Summary statistics} Here, we present a table of the summary statistics of all datasets considered in this study.

\begin{table}[h]
  \caption{Summary statistics of cleaned datasets.}
  \label{tab:statistics}
  \centering
  \scalebox{0.85}
  {
  \begin{tabular}{lccccc}
    \toprule
    \emph{Dataset} & domain & $\lvert V \rvert$ & $ \lvert E \rvert$ & Max. hyperedge size & Number of inferred communities\\
    \midrule
    email-enron & email & 143 & 1,457 & 18 & 5\\
    email-eu & email & 986 & 24,520 & 40 & 42\\
    disgenenet & biology & 1,816 & 7,116 & 20 & 22 \\
    diseasome & biology & 516 & 314 & 11 & N/A\\
    contact-primary-school & proximity & 242 & 12,704 & 5 & 10\\
    contact-high-school & proximity & 327 & 7,818 & 5 & 9\\
    \bottomrule
  \end{tabular}
  }
\end{table}

\newpage

\subsection{email-eu}

The "email-eu" dataset is a higher-order dataset generated from email data sent within a large European research institution between October 2003 and May 2005 (18 months) \cite{landry_xgi-data_2023,leskovec_graph_2007,yin_local_2017,benson_simplicial_2018}. In this dataset, nodes represent email addresses and edges represent email messages. We neither account for the sender-receiver relationships nor the temporality of the emails sent. In addition, we remove duplicate emails involving the same group of people, emails sent to oneself, and email addresses that neither send nor receive emails. This dataset is heterogeneous in both the degree and edge size distributions.

\begin{figure*}[h]
    \centering
    \begin{tabular}{cc}
        \subfloat[Effective information \label{fig:effective_information_email-eu}]{\includegraphics[width=6.75cm]{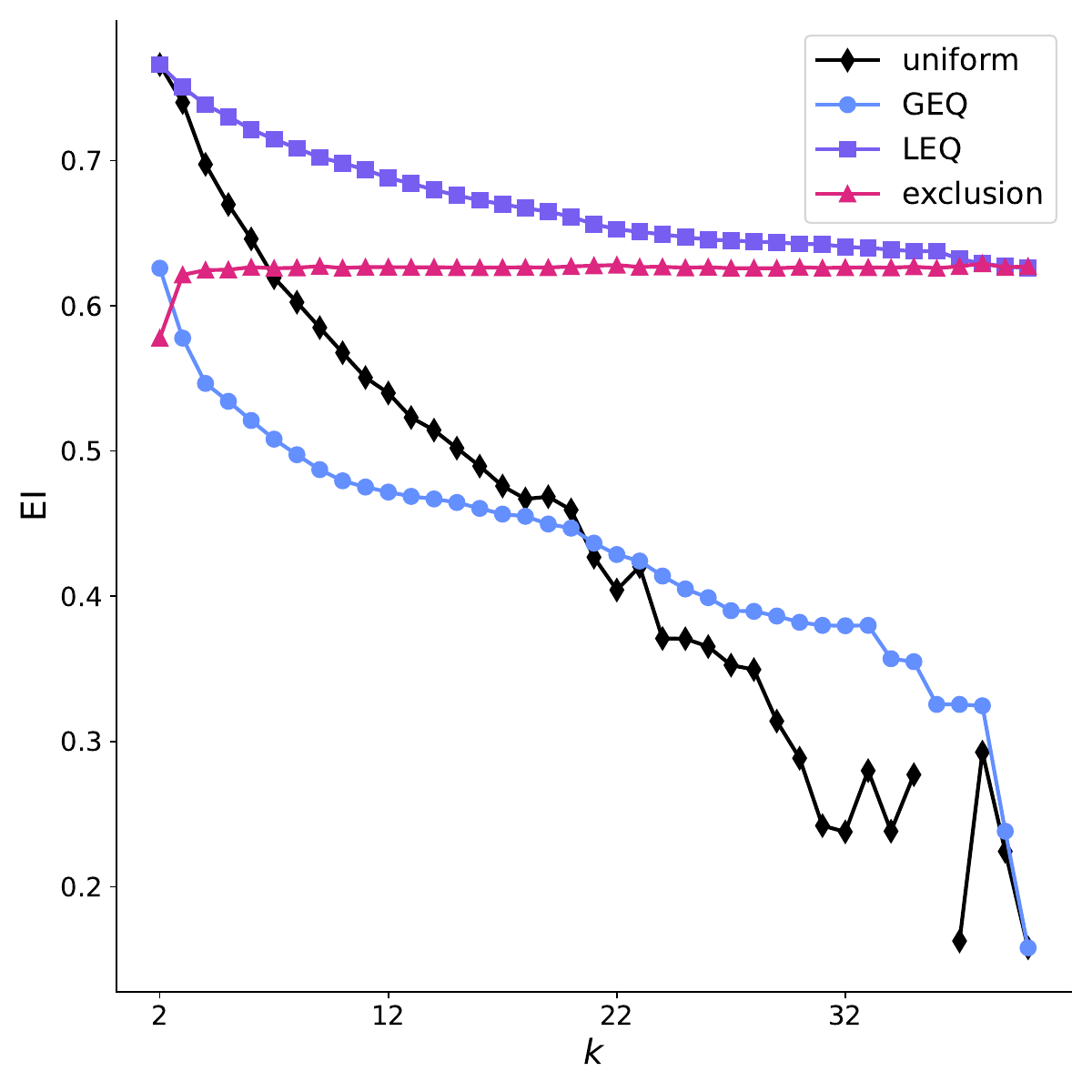}} &  \subfloat[Degree assortativity \label{fig:assortativity_email-eu}]{\includegraphics[width=7cm]{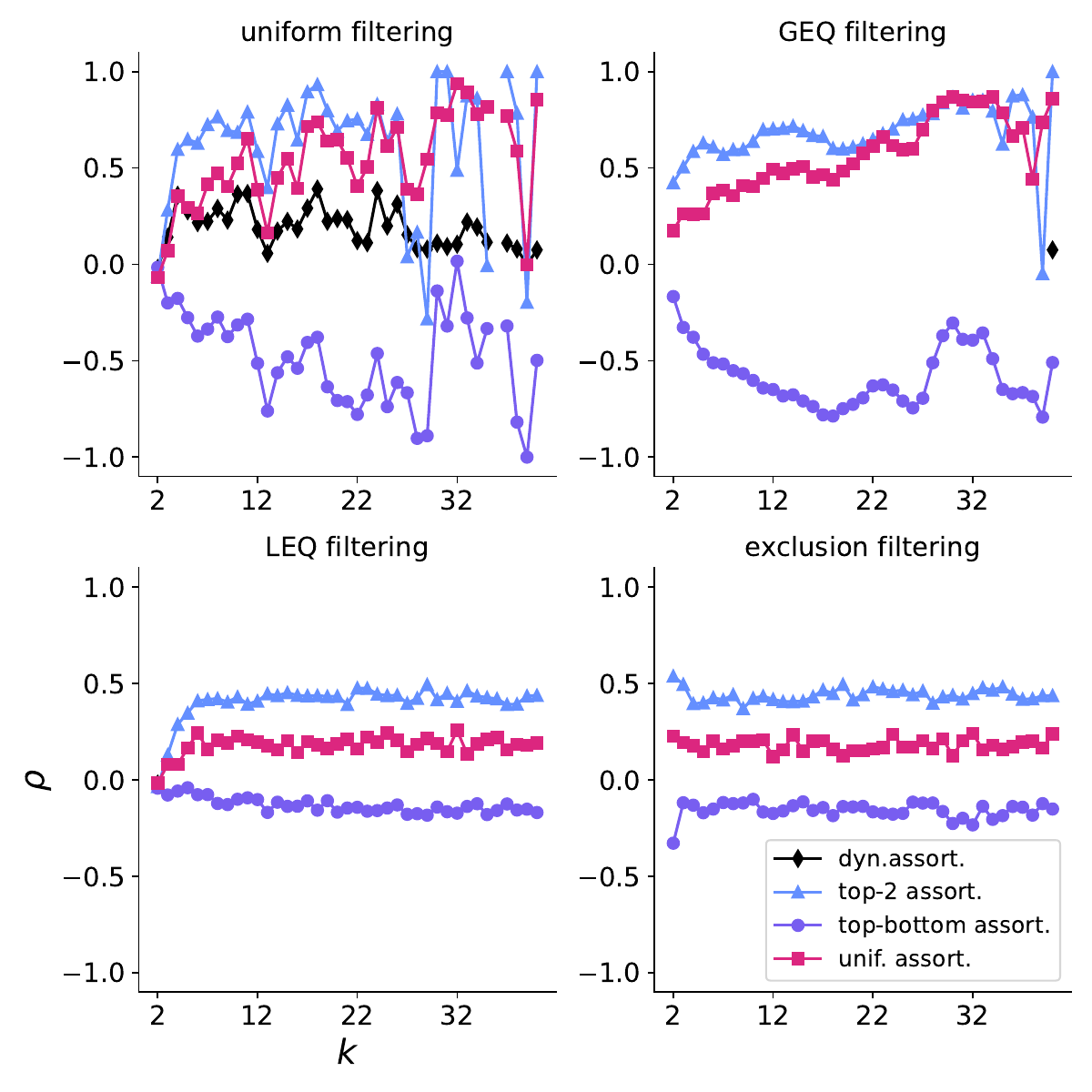}}
    \end{tabular}
    \caption{Global higher-order structural measures}
    \label{fig:global_measures_email-eu}
\end{figure*}
\begin{figure*}[h]
    \centering
    \begin{tabular}{cc}
        \subfloat[$\ell_\infty$-normalized betweenness centrality \label{fig:betweenness_centrality_email-eu}]{\includegraphics[width=7cm]{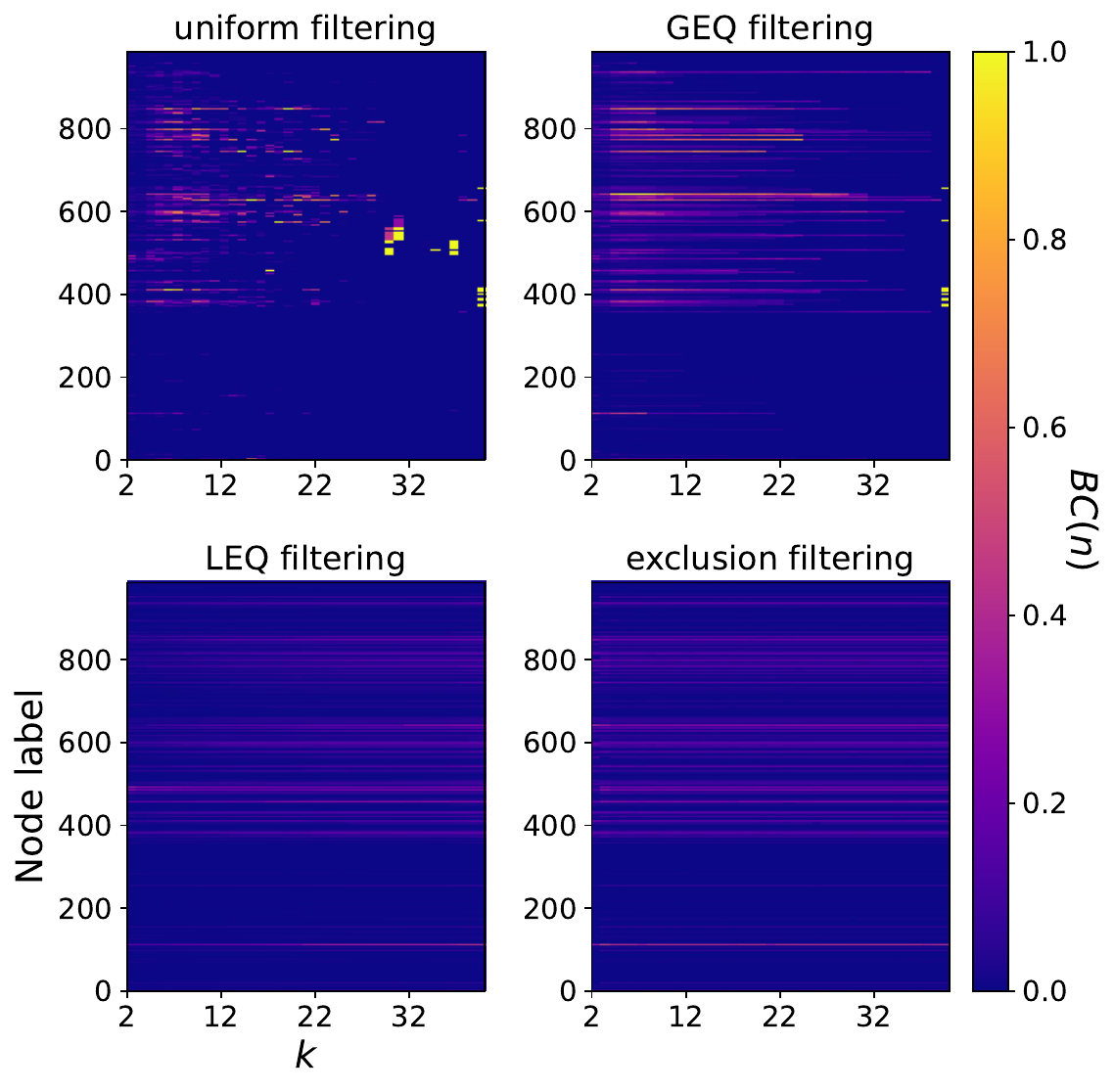}} &  \subfloat[Community labels \label{fig:community_labels_email-eu}]{\includegraphics[width=7cm]{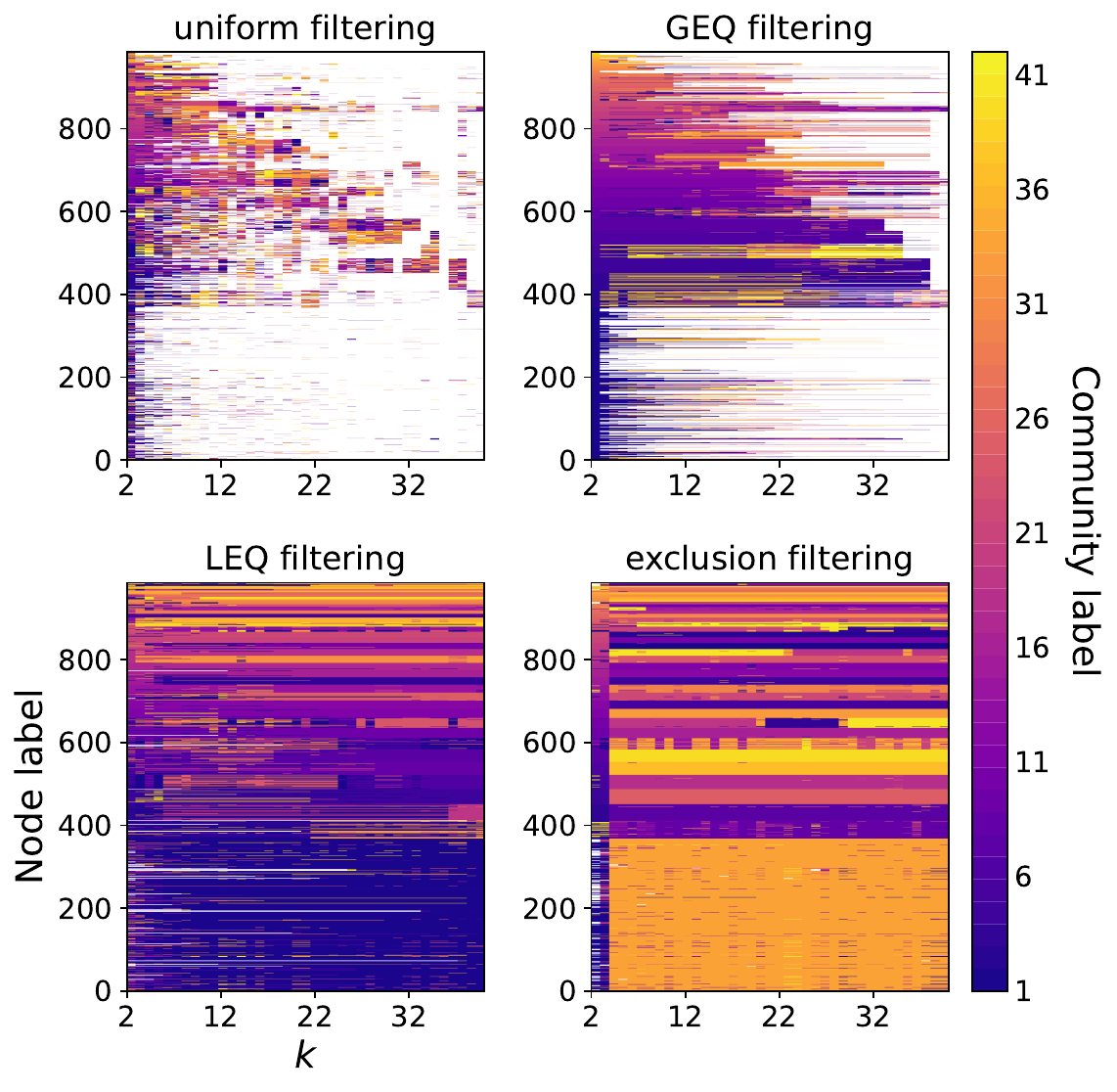}}
    \end{tabular}
    \caption{Nodal higher-order structural measures}
    \label{fig:nodal_measures_email-eu}
\end{figure*}

In \Cref{fig:effective_information_email-eu}, we observe that for the effective information, small interactions contribute the most and that there is a sharp drop-off in effective information for interactions larger than 32 email recipients. We also observe in \Cref{fig:assortativity_email-eu} that as the interaction size increases, the dataset becomes more and more strongly assortative. This is not the case for the top-bottom assortativity, but this could be caused by a highly dense core with a few peripheral nodes so that, on average, the degrees are highly correlated, but when looking at the highest and lowest degrees, they are highly negatively correlated. The community structure is harder to decipher; we have specified 42 communities to infer. Nonetheless, in \Cref{fig:community_labels_email-eu}, we see that as the email size increases, fewer and fewer nodes participate. We also see that excluding interactions of sizes two or three changes the community labels of many nodes. We see several communities persist over larger scales of interactions while other communities are much more sensitive to interaction size. In \Cref{fig:betweenness_centrality_email-eu}, we see that smaller interactions stabilize the centrality, and when excluding larger and larger interactions, the centrality is scale-dependent.

\newpage

\subsection{disgenenet}
The "disgenenet" dataset is a higher-order dataset describing the relationship between genes and the diseases associated with them \cite{pinero_disgenet_2020,landry_xgi-data_2023}. In this dataset, a disease is a node, and a gene is a hyperedge (the dual of the dataset available from XGI-DATA \cite{landry_xgi-data_2023}). We have removed disease-disease correlations to enforce only disease-gene relationships. Before analyzing the dataset, we applied the LEQ filter with $k=20$ for ease of presentation, as the maximum hyperedge size in the complete dataset is in the hundreds. We also excluded diseases with no associated genes and genes that have the same set of associated diseases. 

\begin{figure*}[h]
    \centering
    \begin{tabular}{cc}
        \subfloat[Effective information \label{fig:effective_information_disgenenet}]{\includegraphics[width=6.75cm]{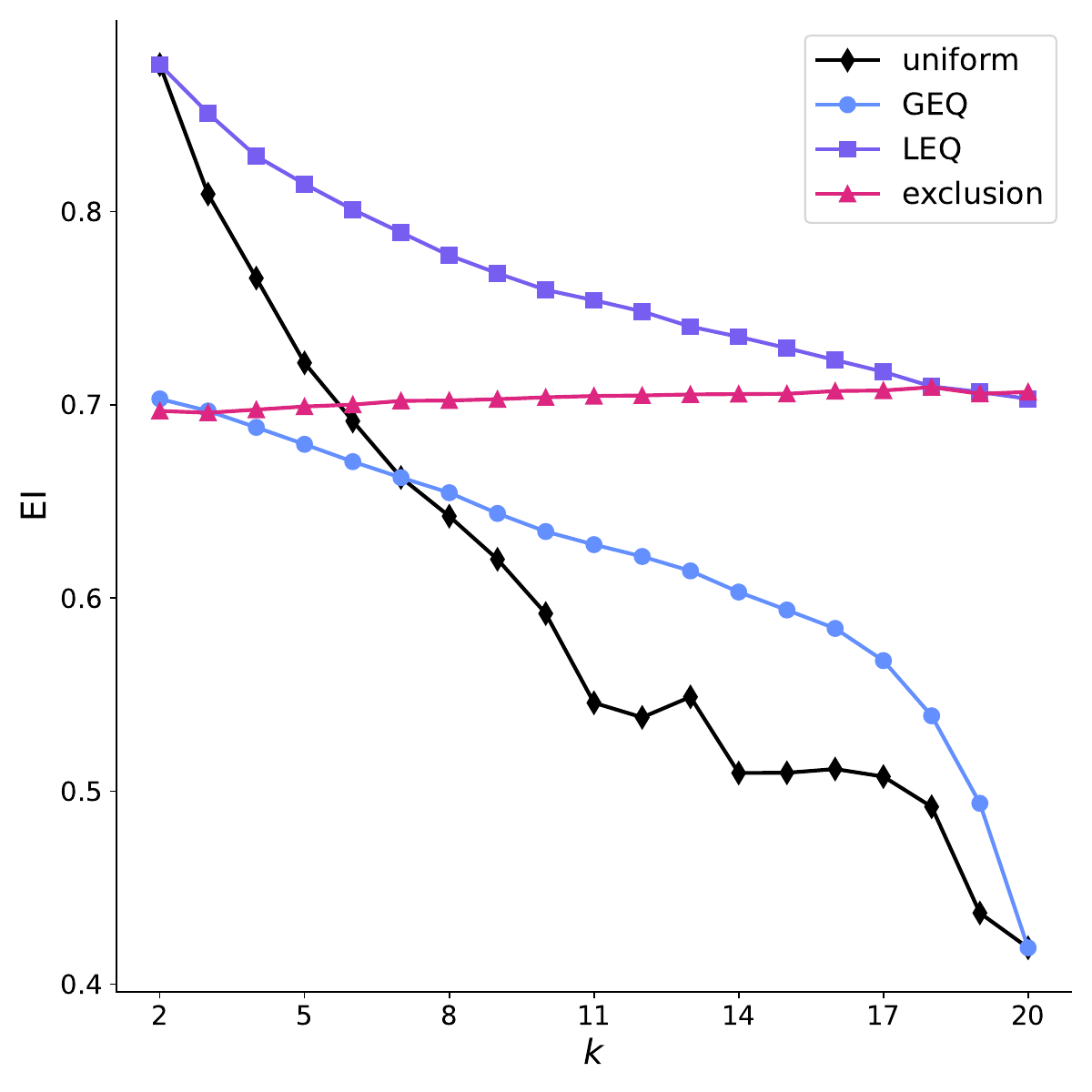}} &  \subfloat[Degree assortativity \label{fig:assortativity_disgenenet}]{\includegraphics[width=7cm]{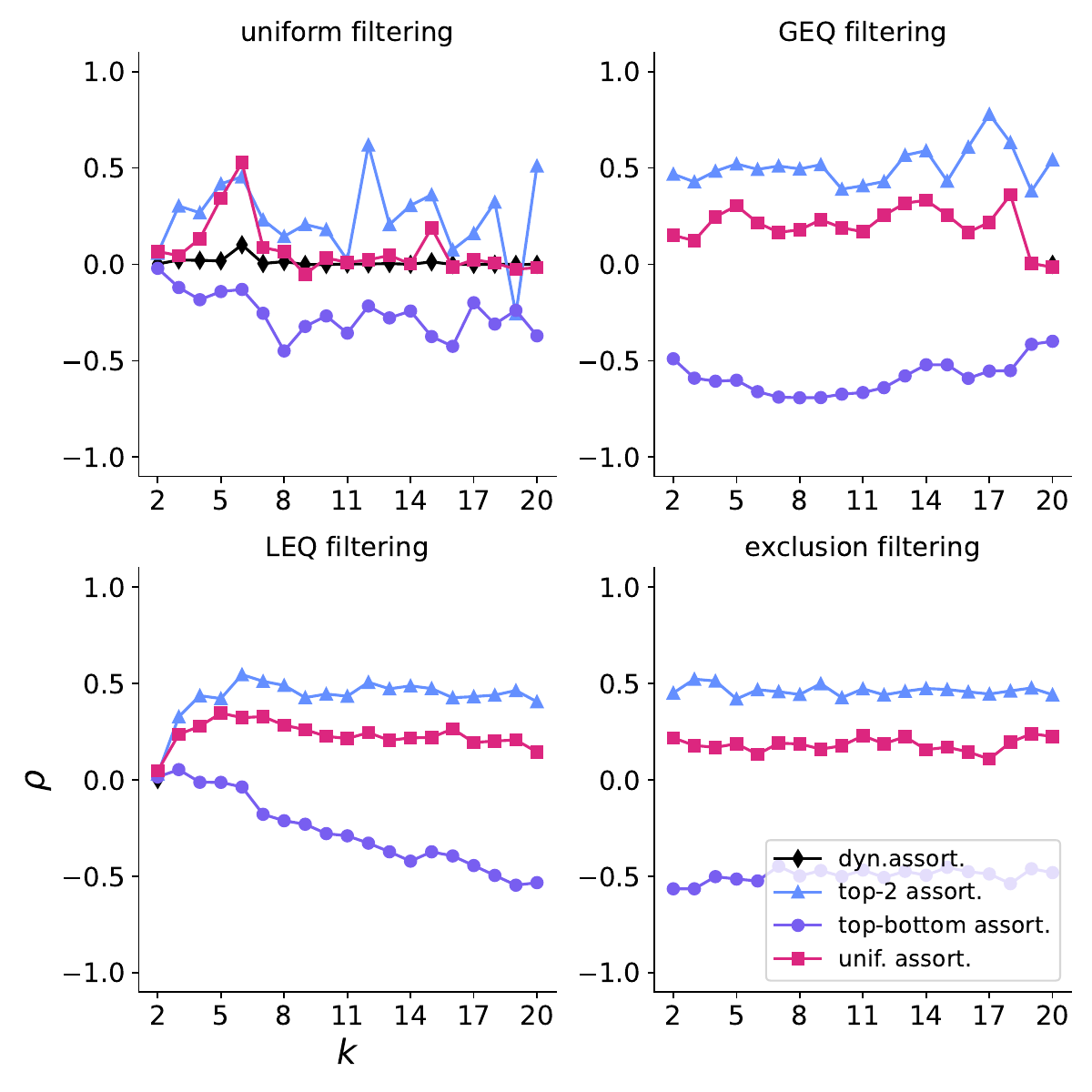}}
    \end{tabular}
    \caption{Global higher-order structural measures}
    \label{fig:global_measures_disgenenet}
\end{figure*}

\begin{figure*}[h]
    \centering
    \begin{tabular}{cc}
        \subfloat[$\ell_\infty$-normalized betweenness centrality \label{fig:betweenness_centrality_disgenenet}]{\includegraphics[width=7cm]{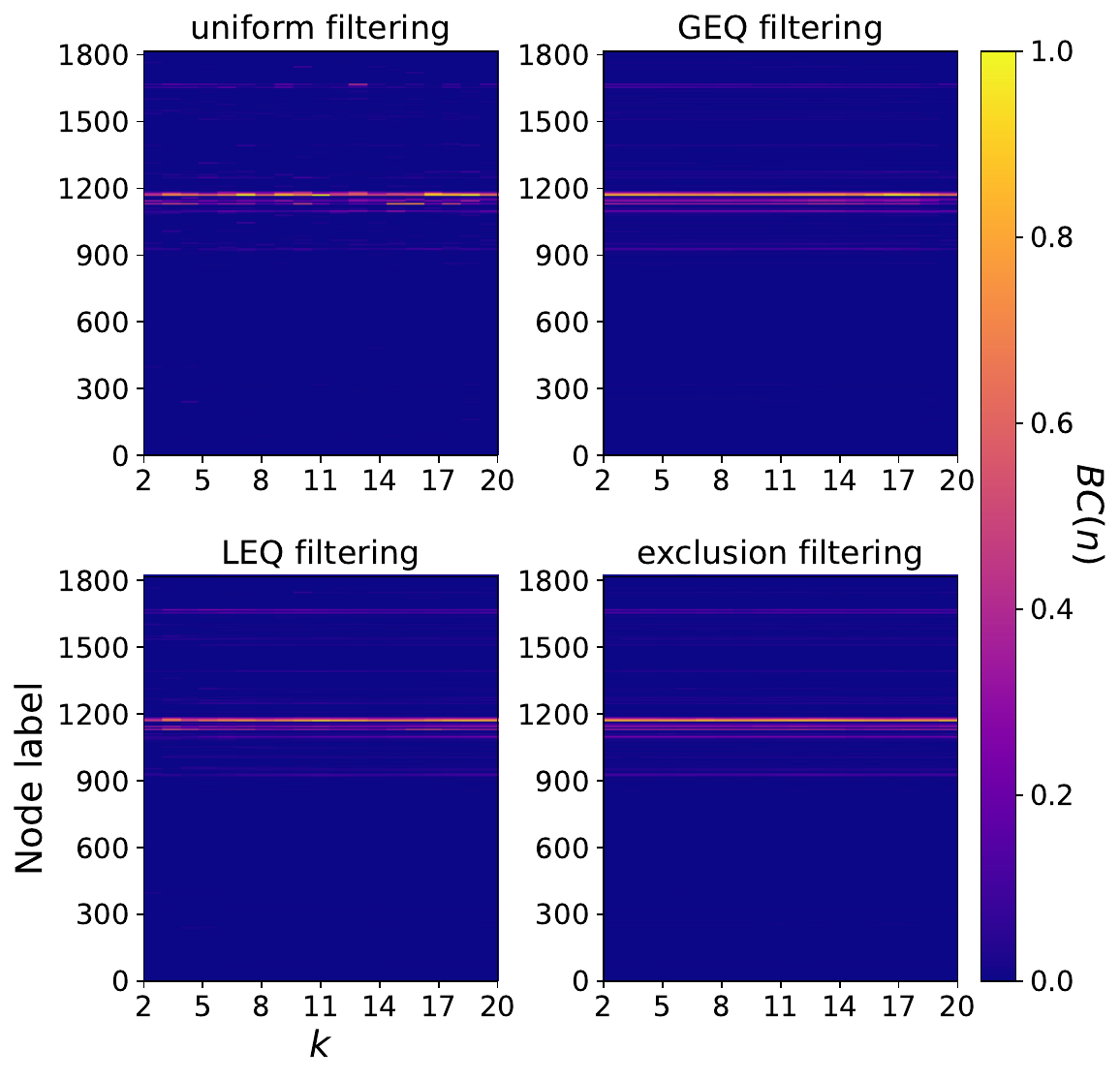}} &  \subfloat[Community labels \label{fig:community_labels_disgenenet}]{\includegraphics[width=7cm]{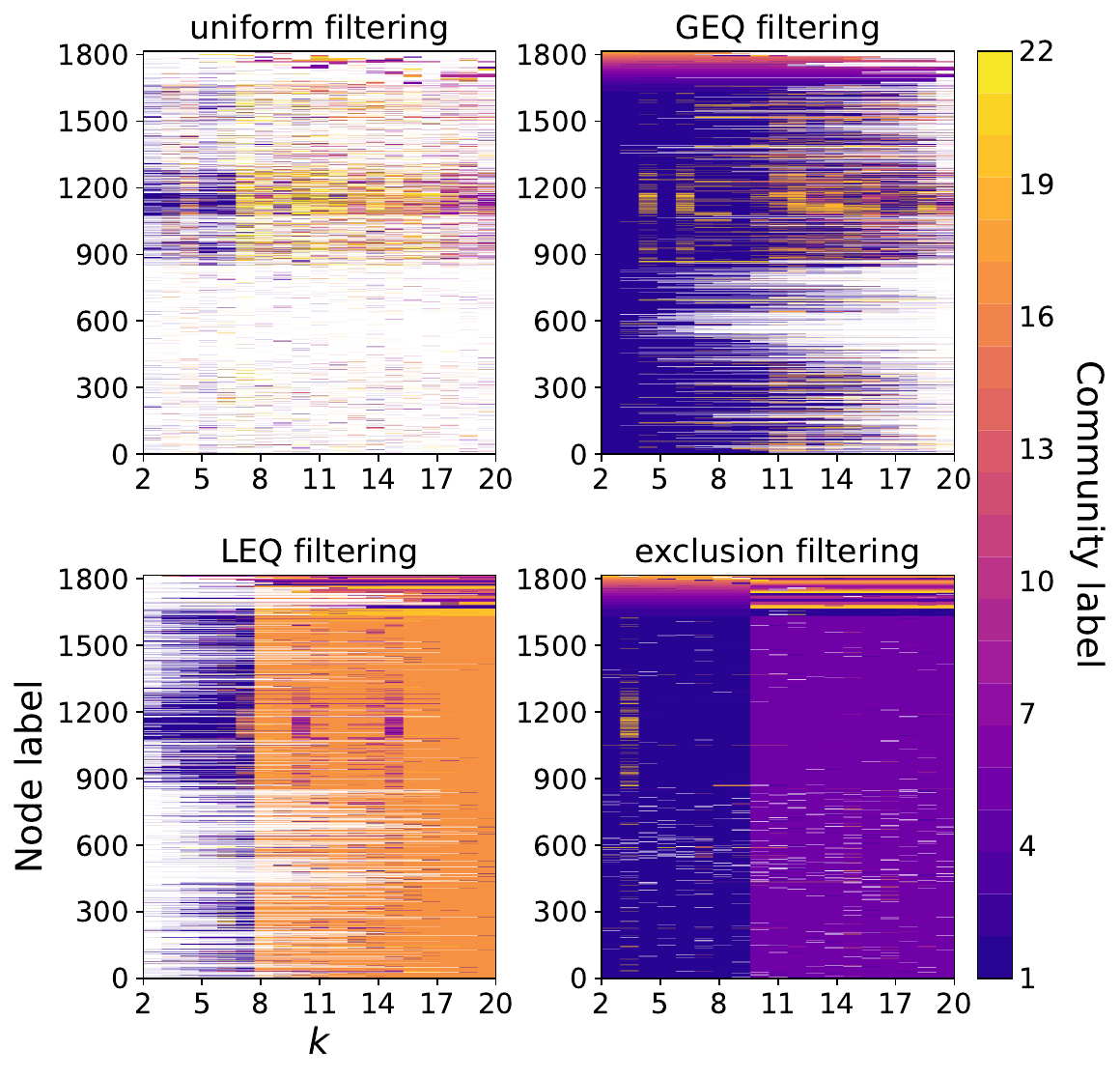}}
    \end{tabular}
    \caption{Nodal higher-order structural measures}
    \label{fig:nodal_measures_disgenenet}
\end{figure*}

In \Cref{fig:effective_information_disgenenet}, the diseasome dataset shows decreasing effective information for increasing interaction size. Unlike the email-eu dataset, the pairwise interactions do not have as significant an influence, and excluding larger and larger interactions slightly increases the effective information. \Cref{fig:assortativity_disgenenet} indicates that associations between diseases become assortative for those involving more than two diseases. The assortativity remains relatively stable, however, and plateaus for mid-sized interactions. As seen in \Cref{fig:betweenness_centrality_disgenenet}, for all filterings, the most central node remains so across most interaction sizes, indicating that the diseases most associated with other diseases are consistent across all scales of interaction. \Cref{fig:community_labels_disgenenet} shows nodes participating in many higher-order interactions, with many nodes not participating in lower-order interactions. While some communities persist across filtering parameters, many nodes often switch communities.

\newpage

\subsection{diseasome}

The "diseasome" dataset tracks diseases and the genes associated with them \cite{goh_human_2007}. In this dataset, a disease is a node, and a gene is a hyperedge. The "label" attribute of the nodes is the disease description, and the "label" attribute of the edges is the gene name. The disease-disease correlations were filtered out to enforce only disease-gene relationships. This results in a very sparse hypergraph, precluding us from analyzing community structure.

\begin{figure*}[h]
    \centering
    \begin{tabular}{cc}
        \subfloat[Effective information \label{fig:effective_information_diseasome}]{\includegraphics[width=7cm]{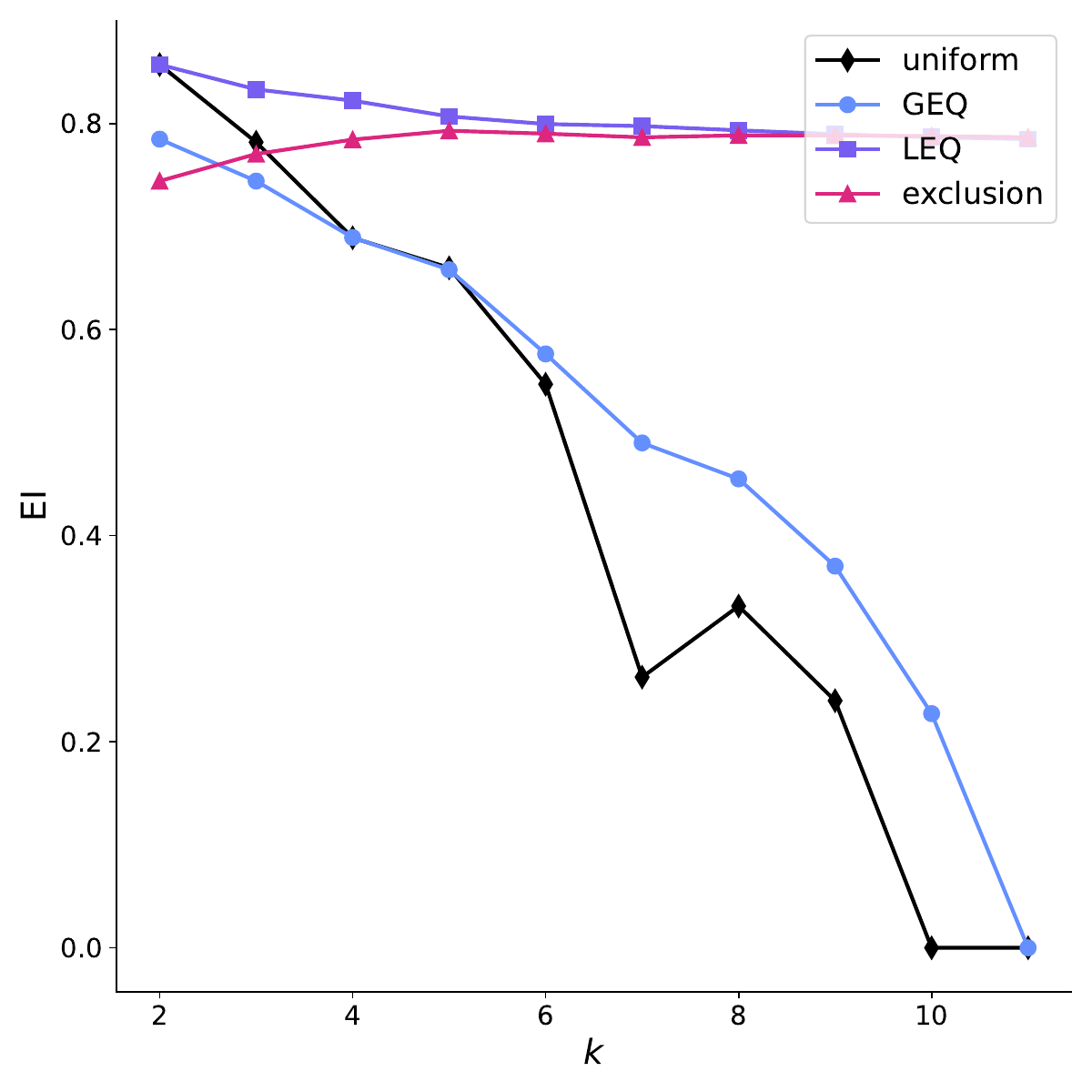}} &  \subfloat[Degree assortativity \label{fig:assortativity_diseasome}]{\includegraphics[width=7cm]{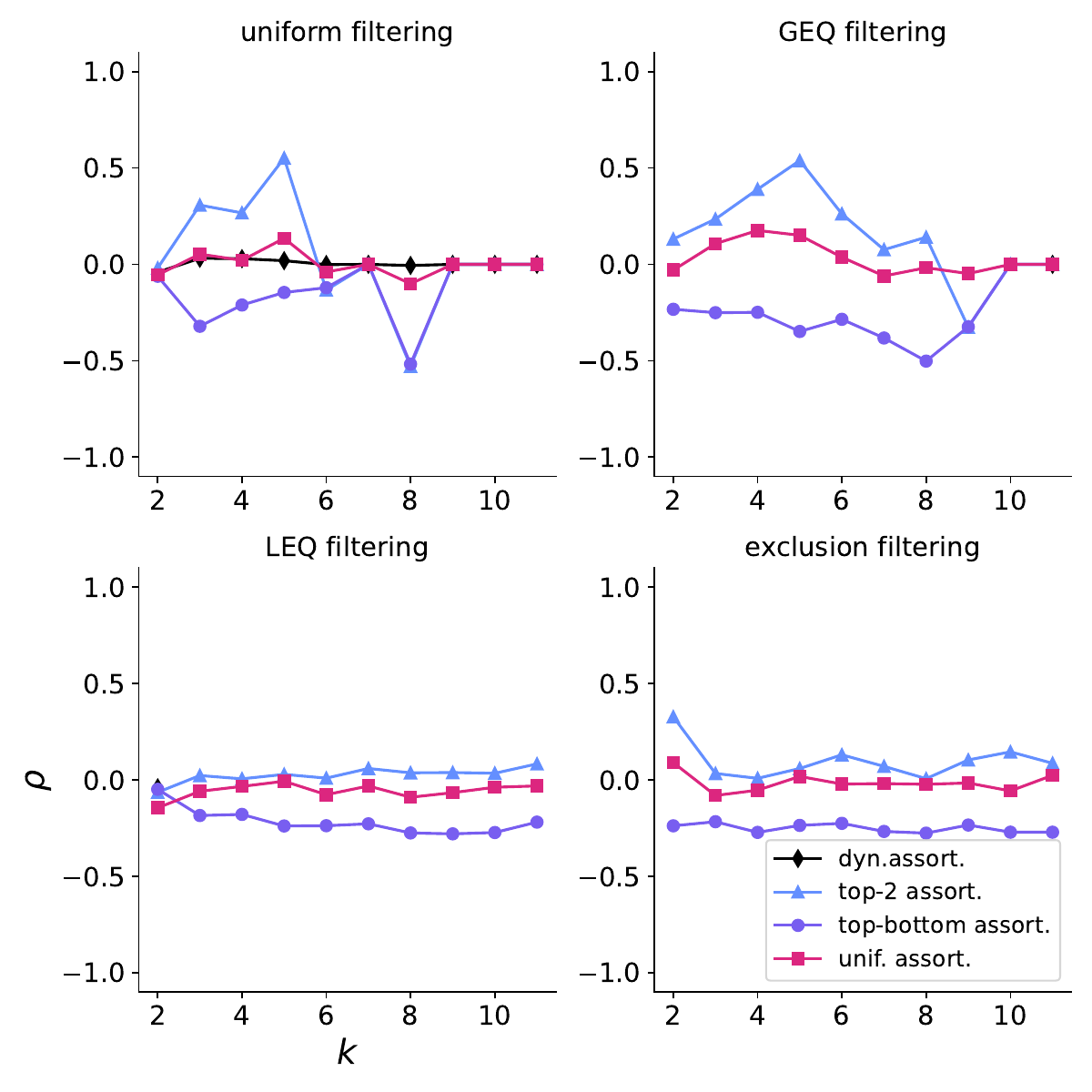}}
    \end{tabular}
    \caption{Global higher-order structural measures}
    \label{fig:global_measures_diseasome}
\end{figure*}

\begin{figure*}[h]
    \centering
    \includegraphics[width=7cm]{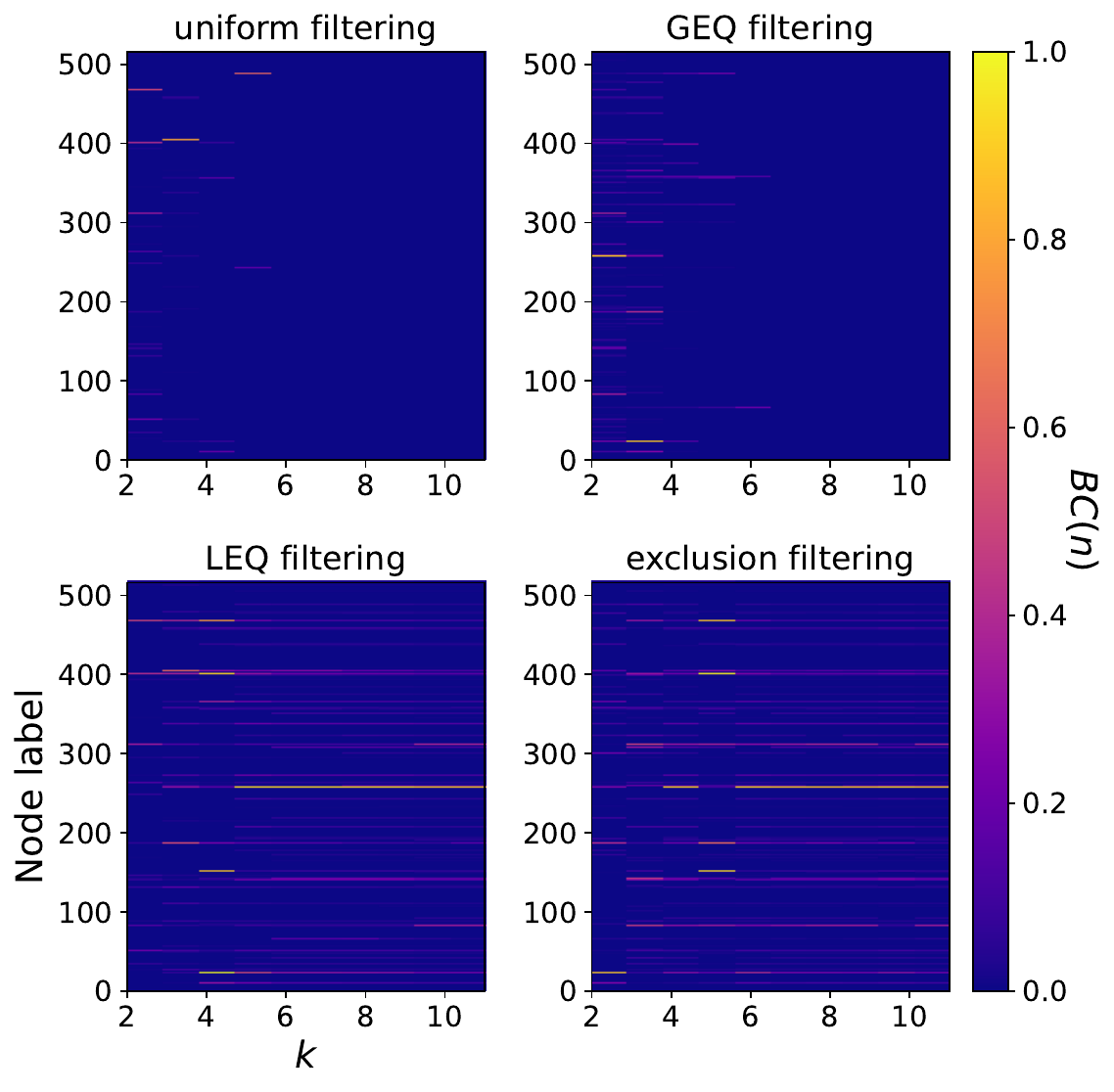}
    \caption{$\ell_\infty$-normalized betweenness centrality}
    \label{fig:diseasome_centrality}
\end{figure*}

When ignoring interactions of smaller sizes through filtering, the effective information (shown in \Cref{fig:effective_information_diseasome}) of the dataset decreases -- although, interestingly, unlike the email datasets, the effective information drops much more sharply for the GEQ filtering, which may be due to the sparsity of larger interactions. This dataset exhibits low assortativity for many filterings and filtering parameters, as seen in \Cref{fig:assortativity_diseasome}. As in other datasets, the pairwise interactions seem to have the most significant effect on the assortativity. As seen from \Cref{fig:diseasome_centrality}, centrality exhibits the most change in the smaller interactions, specifically those smaller than size 6.

\newpage

\subsection{contact-primary-school}
The "contact-primary-school" dataset is a contact network amongst children and teachers at a primary school in Lyon, France. This dataset was collected over two days and inferred via proximity sensing between wearable badges \cite{benson_simplicial_2018,stehle_high-resolution_2011,landry_xgi-data_2023}. We form hyperedges through cliques of simultaneous contacts. Specifically, we construct a hyperedge for every maximal clique amongst the contact edges at each timestamp, and we aggregate all contacts to form a static higher-order network. We remove duplicate contact groups to sparsify the contact network because the dataset is extremely dense.

\begin{figure*}[h]
    \centering
    \begin{tabular}{cc}
        \subfloat[Effective information \label{fig:effective_information_contact-primary-school}]{\includegraphics[width=7cm]{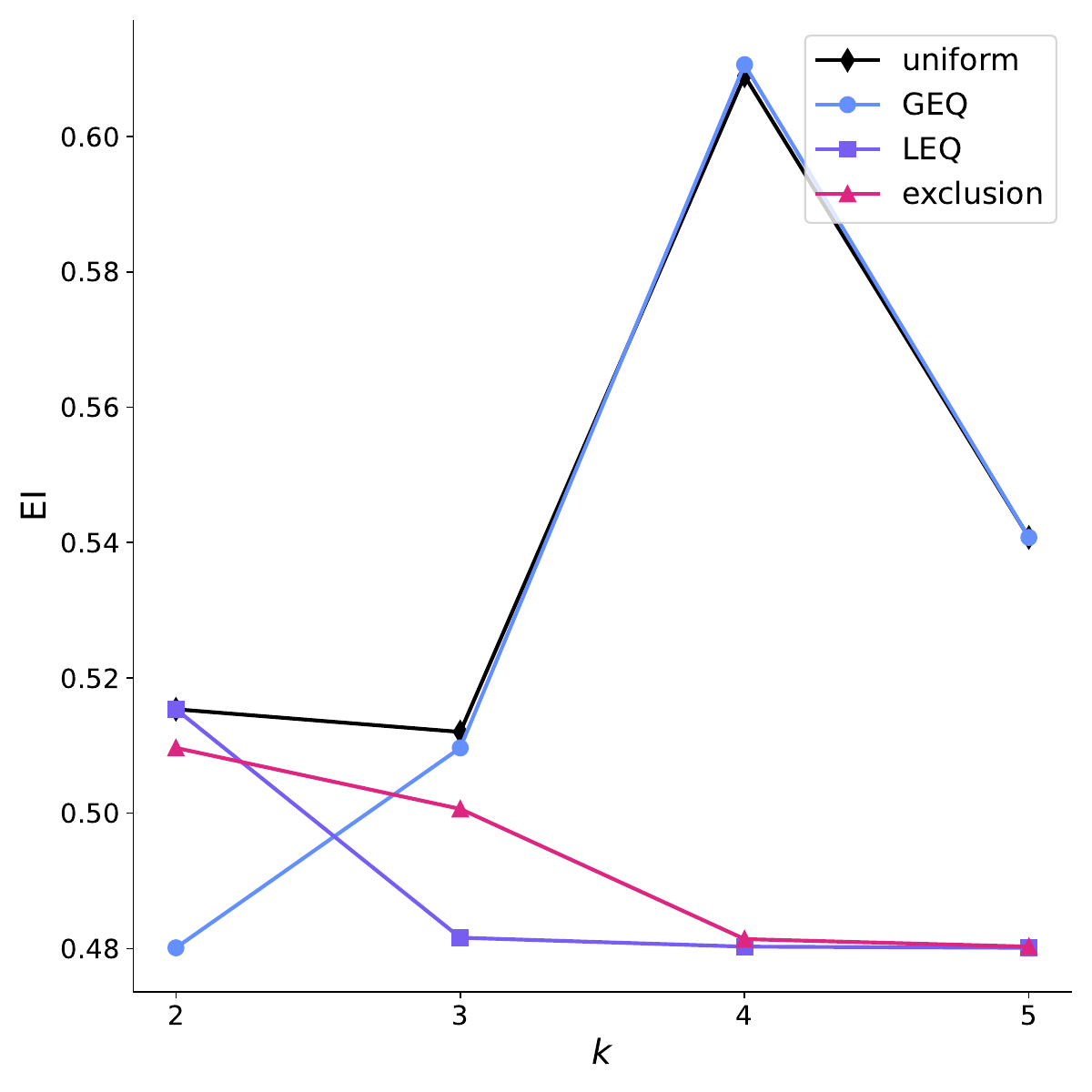}} &  \subfloat[Degree assortativity \label{fig:assortativity_contact-primary-school}]{\includegraphics[width=7cm]{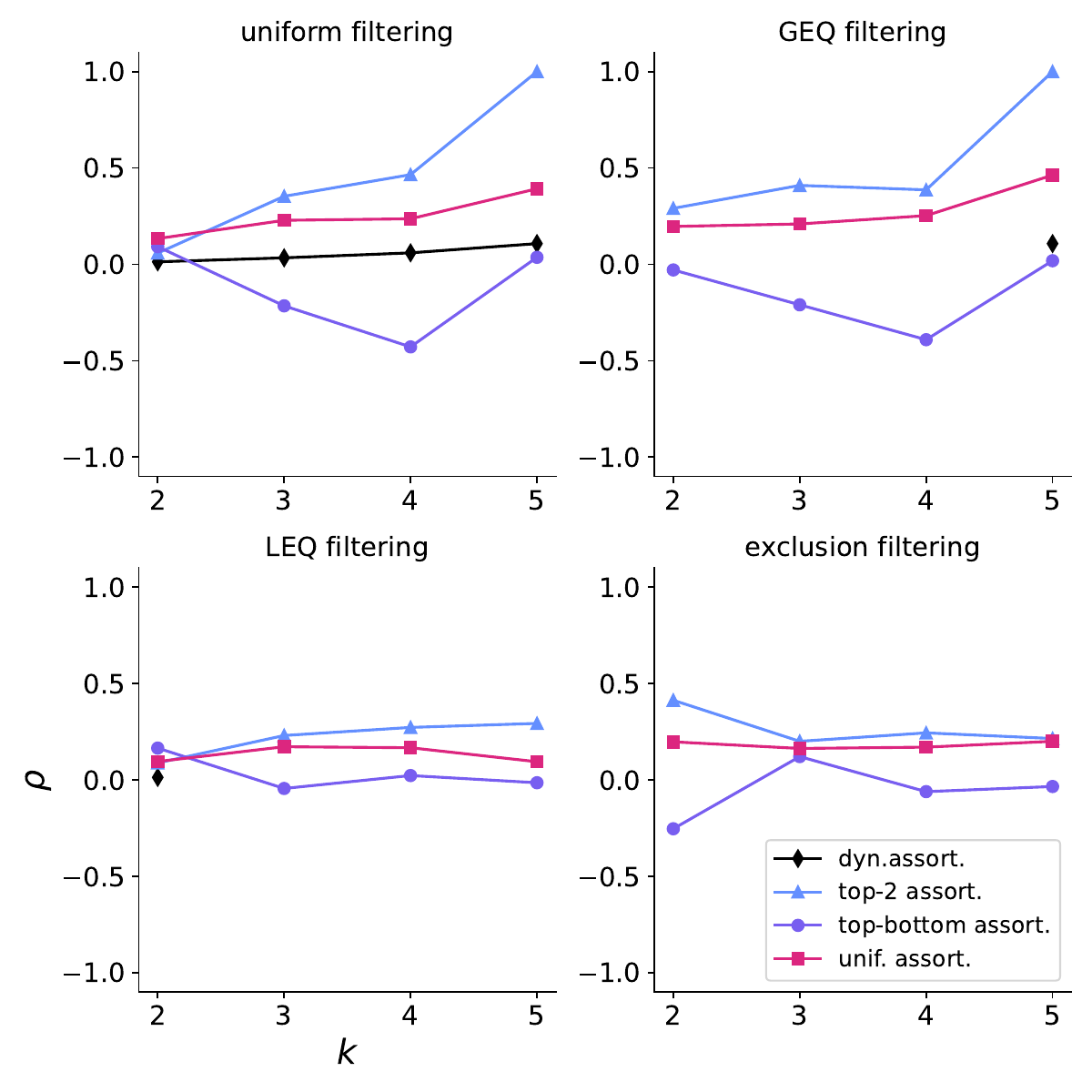}}
    \end{tabular}
    \caption{Global higher-order structural measures}
    \label{fig:global_measures_contact-primary-school}
\end{figure*}

\begin{figure*}[h]
    \centering
    \begin{tabular}{cc}
        \subfloat[$\ell_\infty$-normalized betweenness centrality \label{fig:betweenness_centrality_contact-primary-school}]{\includegraphics[width=7cm]{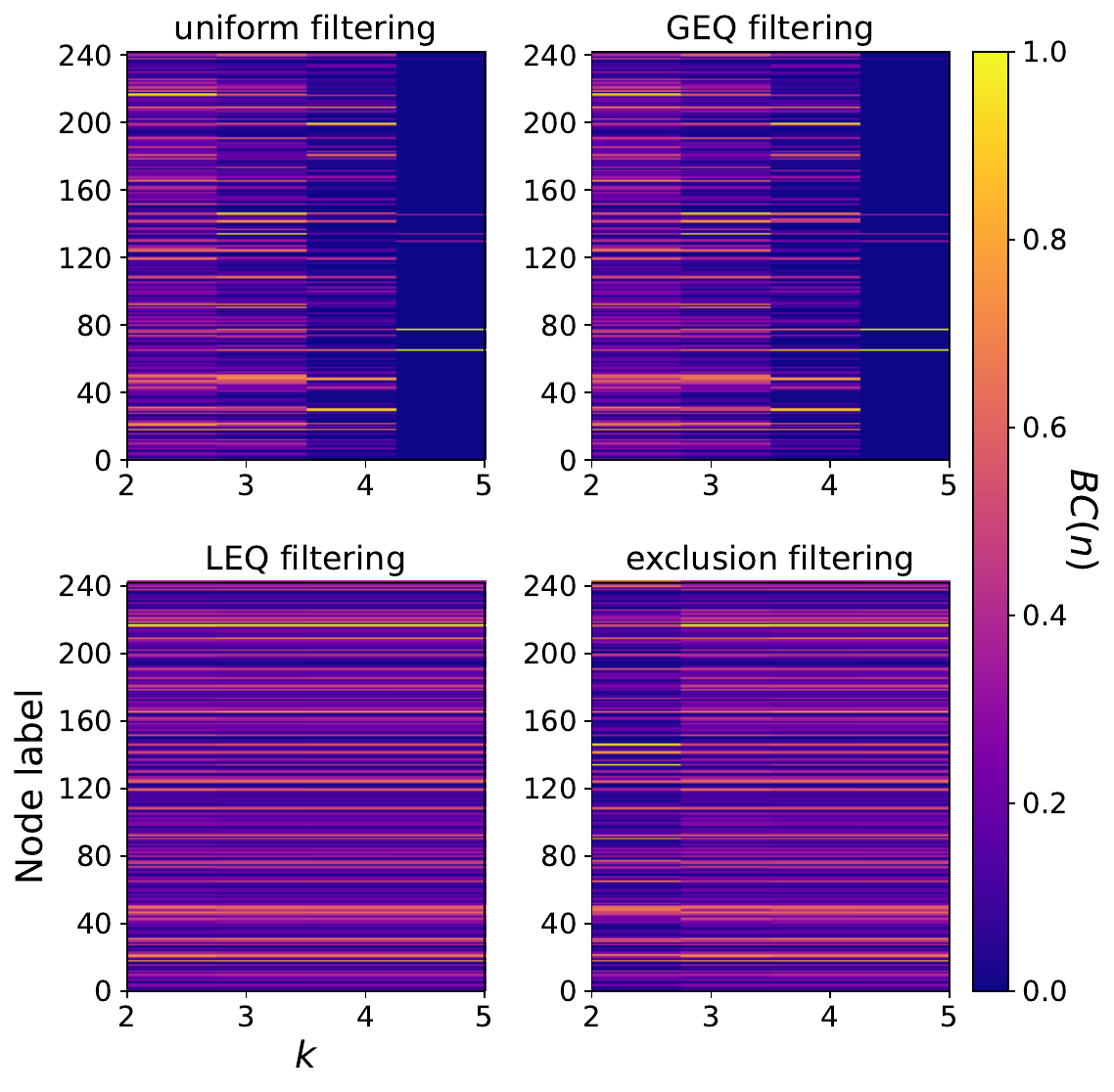}} &  \subfloat[Community labels \label{fig:community_labels_contact-primary-school}]{\includegraphics[width=7cm]{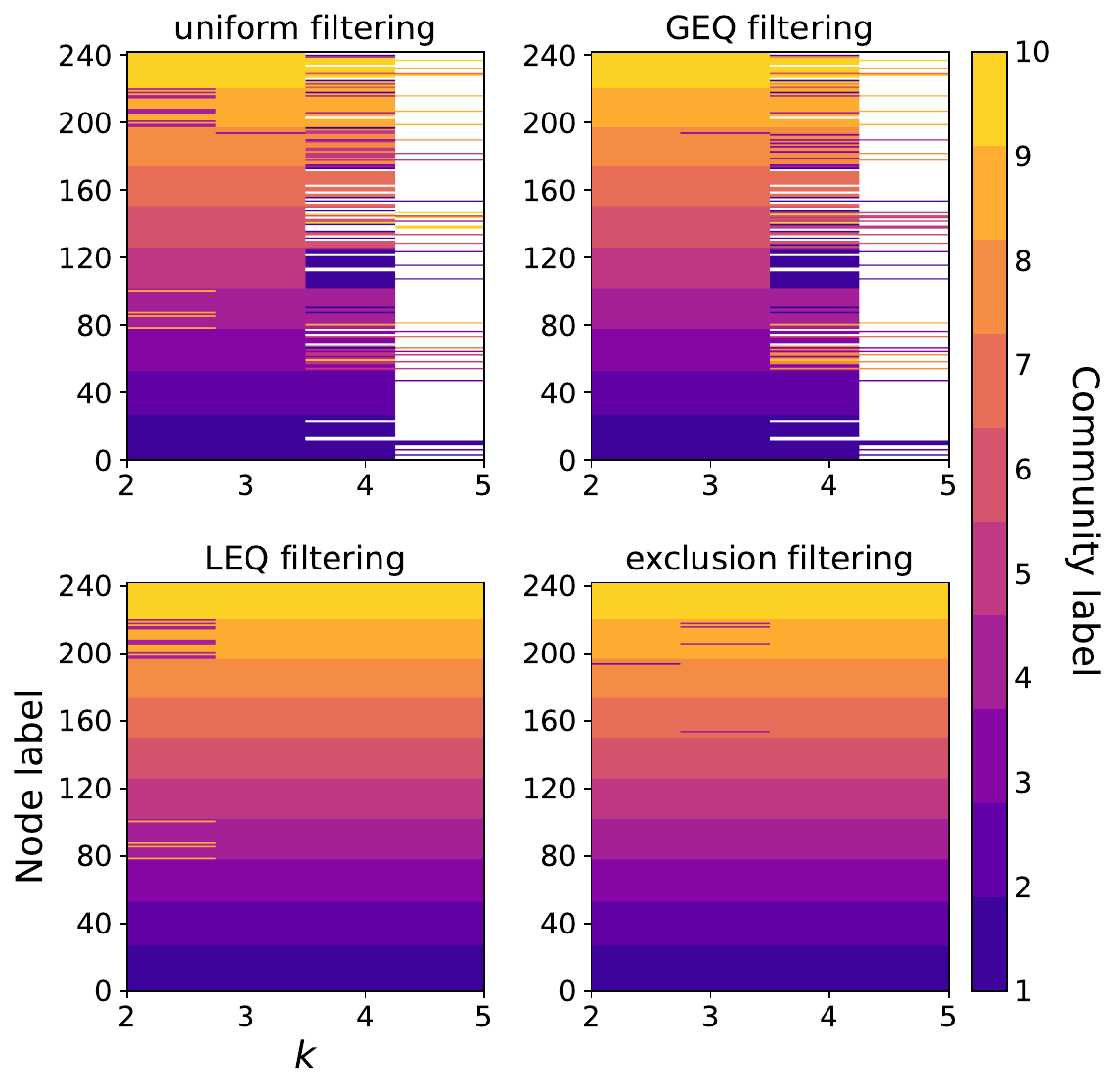}}
    \end{tabular}
    \caption{Nodal higher-order structural measures}
    \label{fig:nodal_measures_contact-primary-school}
\end{figure*}

Proximity datasets can often have smaller ranges of interaction sizes not only because of physical limitations (in this study, proximity sensors detect interactions in the range of 1-1.5 meters \cite{stehle_high-resolution_2011}), but also because the method of reconstruction requires a complete clique to infer a hyperedge. We note that in \Cref{fig:effective_information_contact-primary-school}, in contrast to the preceding datasets, the effective information is \textit{highest} for larger interaction sizes. The assortativity shown in (\Cref{fig:assortativity_contact-primary-school}) seems to --- on the whole --- gradually increase with interaction si e. The most significant change is in the top-bottom assortativity, where interactions of size 4 are most disassortative. This indicates that the lowest and highest degrees are more anti-correlated, i.e., large-degree nodes tend not to interact with small-degree nodes. We see that the most central nodes are sensitive to pairwise interactions and that when excluding pairwise interactions, the centrality can change quite randomly with interaction size. We see that the community structure is relatively stable; if anything, \Cref{fig:community_labels_contact-primary-school} indicates that very few nodes participate in interactions larger than 4. This could be an artifact of the proximity data collection method.

\newpage

\subsection{contact-high-school}
This is a temporal hypergraph dataset, which here means a sequence of timestamped hyperedges where each hyperedge is a set of nodes. This dataset is constructed from a contact network amongst high school students in Marseilles, France, in December 2013, inferred via wearable badges \cite{benson_simplicial_2018,mastrandrea_contact_2015,landry_xgi-data_2023}. We form hyperedges through cliques of simultaneous contacts. Specifically, for every unique timestamp in the dataset, we construct a hyperedge for every maximal clique amongst the contact edges that exist for that timestamp. Timestamps were recorded in 20-second intervals, and we aggregated these contacts to form a static higher-order network. Because of the high density of this data, we remove duplicate interactions.

\begin{figure*}[h]
    \centering
    \begin{tabular}{cc}
        \subfloat[Effective information \label{fig:effective_information_contact-high-school}]{\includegraphics[width=7cm]{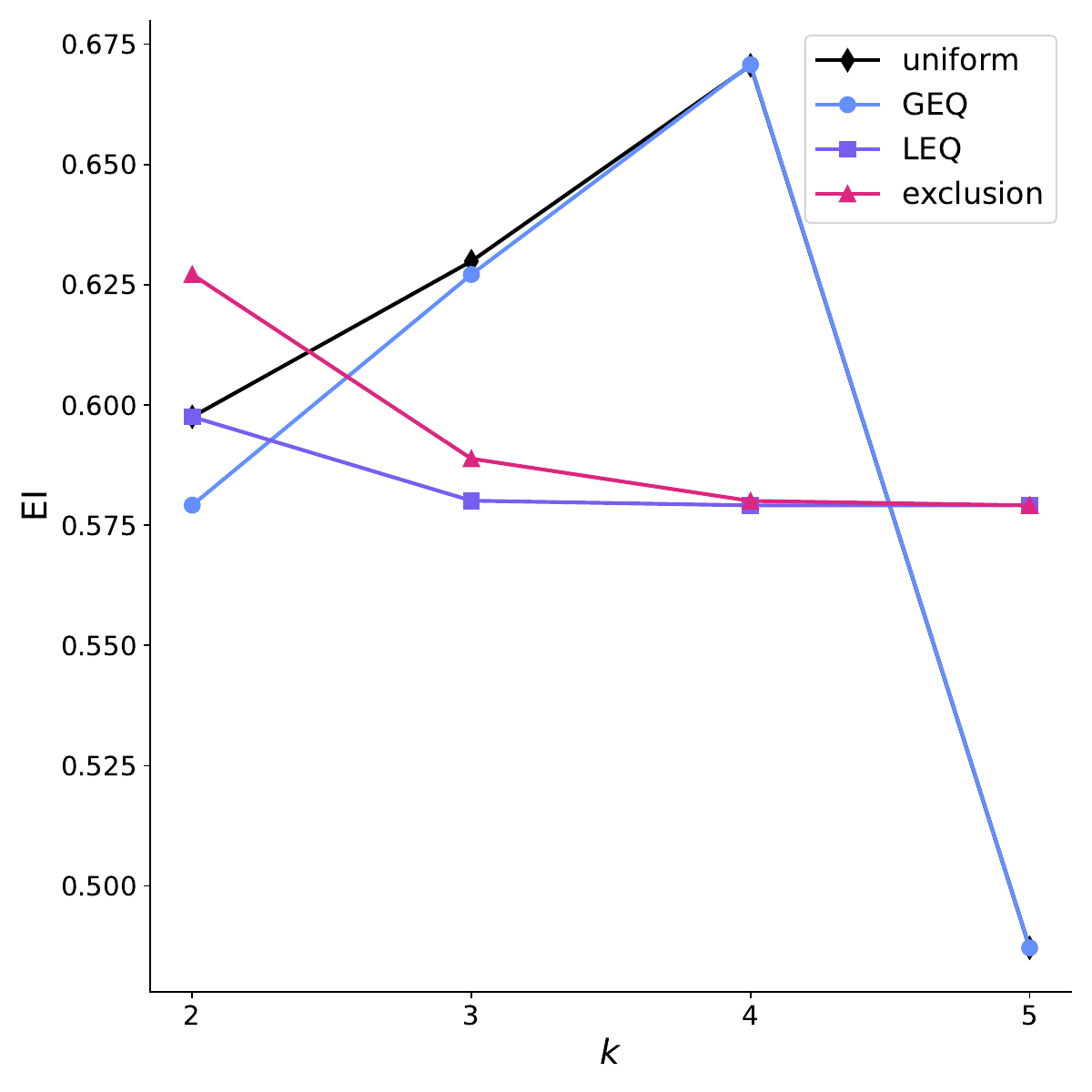}} &  \subfloat[Degree assortativity \label{fig:assortativity_contact-high-school}]{\includegraphics[width=7cm]{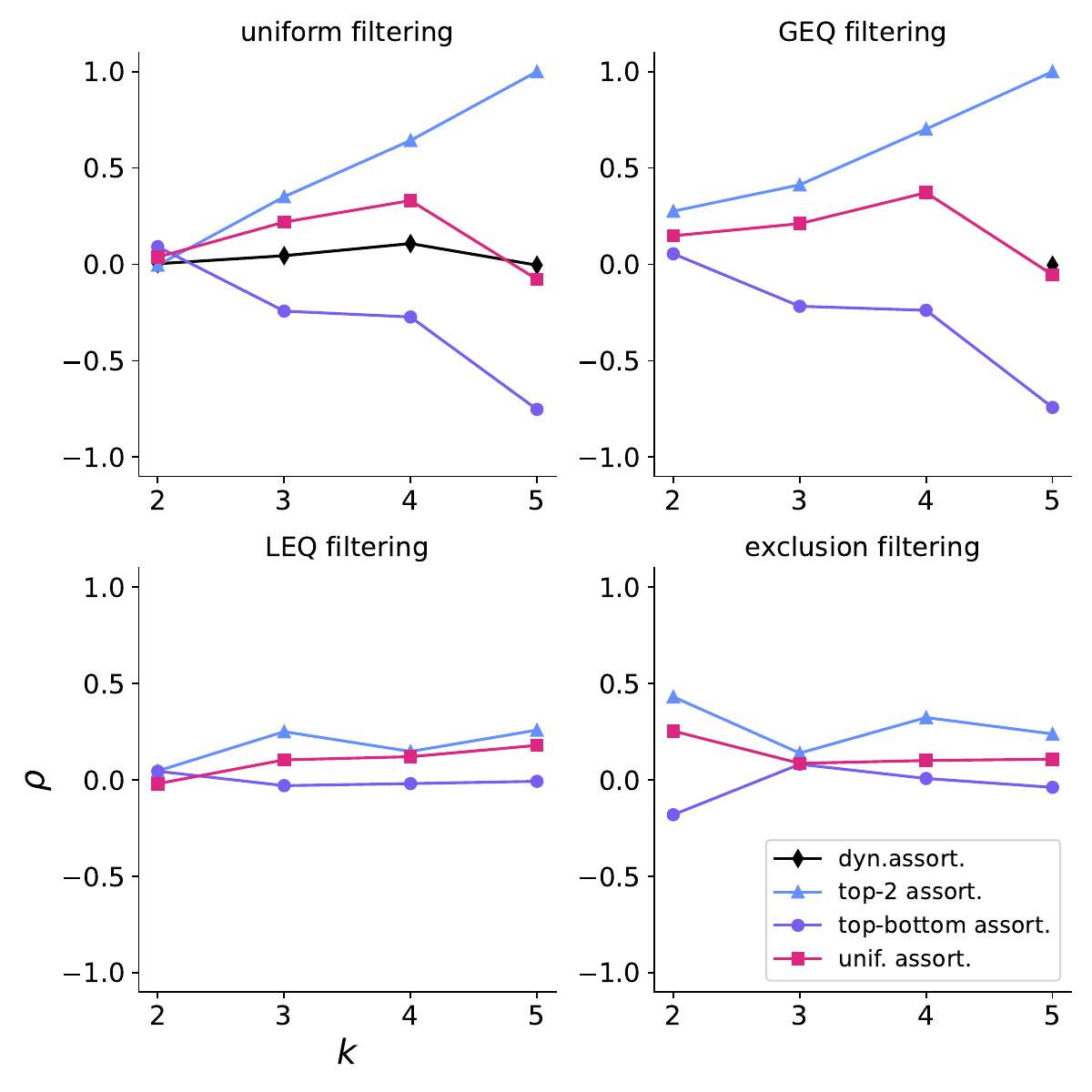}}
    \end{tabular}
    \caption{Global higher-order structural measures}
    \label{fig:global_measures_contact-high-school}
\end{figure*}

\begin{figure*}[h]
    \centering
    \begin{tabular}{cc}
        \subfloat[$\ell_\infty$-normalized betweenness centrality \label{fig:betweenness_centrality_contact-high-school}]{\includegraphics[width=7cm]{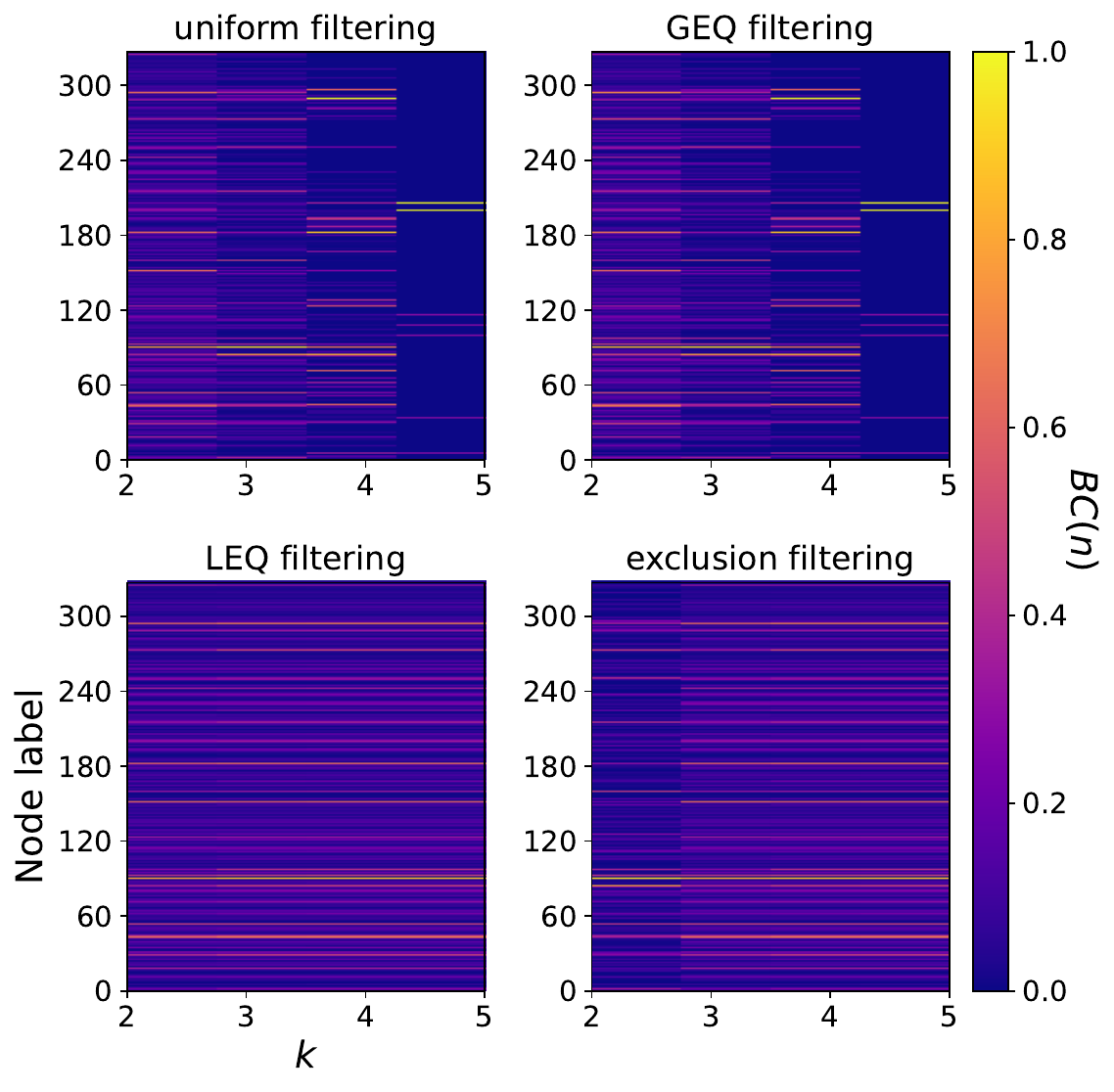}} &  \subfloat[Community labels \label{fig:community_labels_contact-high-school}]{\includegraphics[width=7cm]{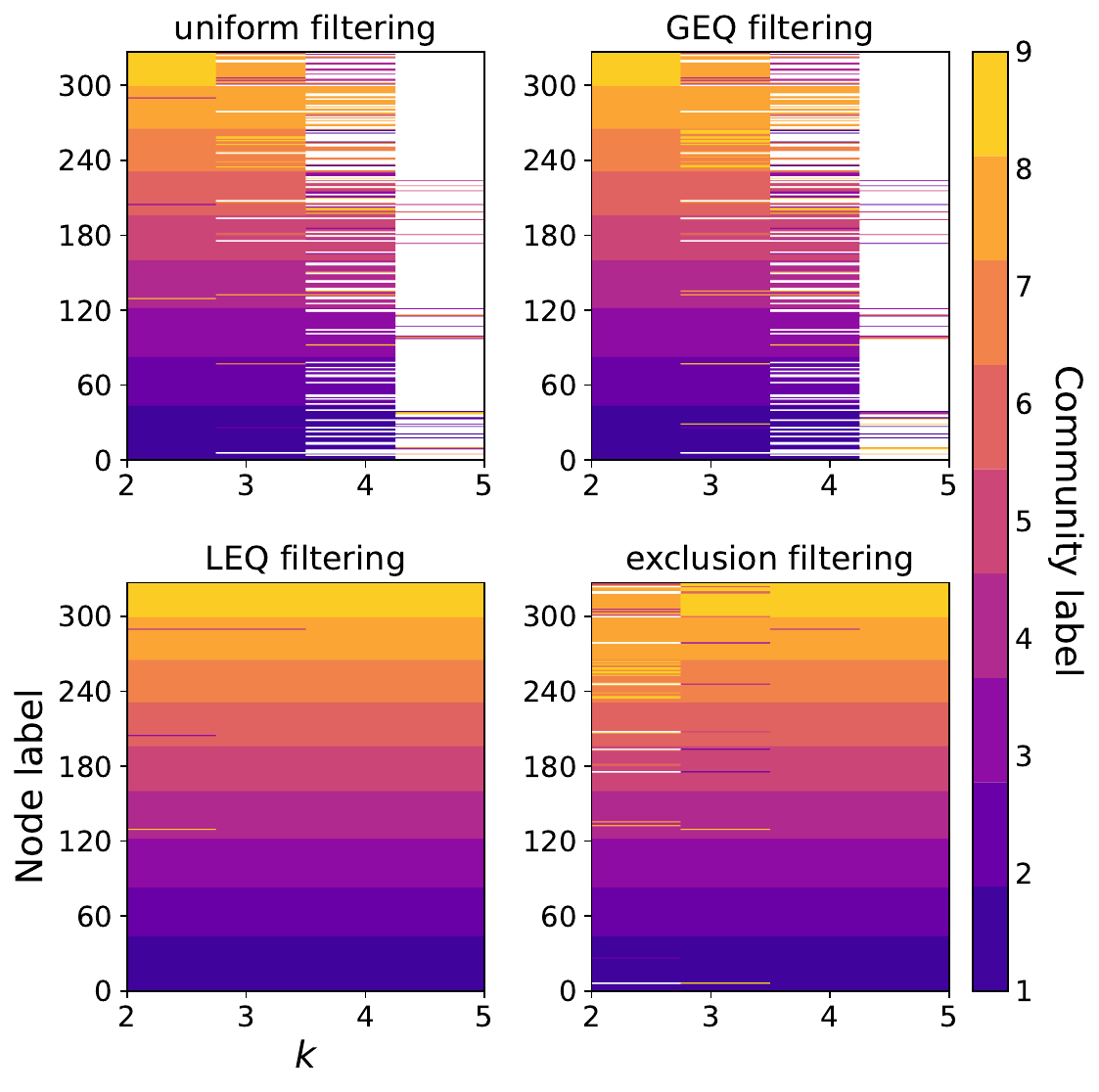}}
    \end{tabular}
    \caption{Nodal higher-order structural measures}
    \label{fig:nodal_measures_contact-high-school}
\end{figure*}

This dataset has similar limitations and artifacts due to the data collection method used for the contact-primary-school dataset. Similar to that dataset, interactions of size 4 have the most effective information, but in this case, interactions of size 5 have the least effective information. The assortativity, centrality, and community structure yield similar conclusions to those of the contact-primary-school dataset.


\section{Domain data insights}

Analyzing email, biological, and proximity datasets allows us to see patterns in structural measures --- effective information, assortativity, centrality, and community structure --- across different domains.

In the email and biological datasets, effective information in the uniform, GEQ, and LEQ filterings largely trends downward with the filtering parameter, while for the exclusion filtering, it largely trends upward. This indicates that higher-order interactions lead to fewer unique associations within these domains. However, in the proximity datasets, we observe the effective information trending upward in the uniform and GEQ filterings (except for $k=5$ in contact-high-school) and downward in the LEQ and exclusion filterings. This indicates the presence of more unique associations among higher-order interactions in this domain. Furthermore, the plots within domains for email and proximity datasets look remarkably similar. However, the plots within the biological datasets look markedly different, which we believe to be an artifact of the large disparity in density.

For the email datasets, the assortative structure seems to come from lower-order interactions and increases in assortativity up to intermediate interaction sizes. The effective information is most affected by the low-order interactions and generally decreases as the filtering parameter is increased.
For the biological datasets, they seem to be neither assortative nor disassortative. The change in the top-bottom or top-2 assortativity measures could be because of changes in the degree structure. For the proximity datasets, in all but the exclusion filtering, we generally observe more assortative interactions as $k$ increases. In general, across all domains, top-bottom assortativity tends to decrease with $k$, except for the exclusion filtering, in which it decreases -- though in all cases, the measure indicates disassortativity.

Across most\footnote{The outlier is disgenenet, where higher-order interactions are prominent, leading to stability within all filterings.} datasets, the betweenness centrality within LEQ and exclusion filterings is more stable than within uniform and GEQ filterings. This stability stems from the fact that among most datasets studied, lower-order interactions are significantly more numerous than higher-order interactions. Nonetheless, the different filterings offer complementary information and deep data insights about each dataset on the node level, such as the ones presented in the main text for the email-enron dataset.

The community structure is clear-cut in the proximity datasets, as communities correspond to students' classrooms and persists regardless of the filtering. The community structure is less evident in the email datasets -- especially in the case of email-eu, where there are 42 ground truth labels. In this domain, we see that community structure is significantly impacted by interaction size, illustrating the discriminatory power of our approach. We refrain from commenting on community structure trends in the biological datasets since sparsity of diseasome precludes community structure analysis.
\twocolumngrid
\bibliography{references}
\end{document}